\documentclass{article}

\usepackage{arxiv}

\usepackage[utf8]{inputenc} 
\usepackage[T1]{fontenc}    
\usepackage{hyperref}       
\usepackage{url}            
\usepackage{booktabs}       
\usepackage{array}
\usepackage{amsfonts}       
\usepackage{nicefrac}       
\usepackage{microtype}      
\usepackage{lipsum}		
\usepackage{amsthm}
\usepackage{graphicx}
\usepackage{doi}
\usepackage{physics}
\usepackage{xcolor}
\usepackage{algorithm}
\usepackage{algpseudocode}
\usepackage{amssymb}
\usepackage{pifont}

\usepackage{setspace}
\title{Slice-to-volume reconstruction with Deep Ritz regularization and stochastic integration}

\date{} 					

\author{
    Conor Rowan \\
	Smead Aerospace Engineering Sciences\\
    3775 Discovery Drive \\
	University of Colorado Boulder\\
	Boulder, CO 80309 \\
	\texttt{conor.rowan@colorado.edu} \\
    \And
    Sumedh Soman\\
	Smead Aerospace Engineering Sciences\\
    3775 Discovery Drive \\
	University of Colorado Boulder\\
	Boulder, CO 80309 \\
	\texttt{sumedh.soman@colorado.edu}
    \And
    John Evans\\
	Smead Aerospace Engineering Sciences\\
    3775 Discovery Drive \\
	University of Colorado Boulder\\
	Boulder, CO 80309 \\
	\texttt{john.evans@colorado.edu} \\
}

\renewcommand{\headeright}{}
\renewcommand{\undertitle}{}
\renewcommand{\shorttitle}{}

\hypersetup{
    pdftitle={Slice-to-volume reconstruction with Deep Ritz regularization and stochastic integration},
    pdfauthor={Conor Rowan, Sumedh Soman, John Evans},
    pdfkeywords={Slice-to-volume reconstruction, Physics-informed machine learning, Deep Ritz method, Inverse problems}
}

\begin{document}
\maketitle

\begin{abstract}
    We present a training-data-free method for reconstructing three-dimensional solids from sparse, noisy planar cross-sections. Sparse cross-sections fundamentally underdetermine the geometry between observed planes, so every reconstruction method supplies a prior. Rather than learn that prior from an offline dataset or identify the reconstruction solely with a minimizer of a continuum
    regularization functional, we study a prior induced jointly by an anisotropic phase-field energy, a neural implicit representation, stochastic underintegration, and gradient-based optimization. The phase-field objective includes a data term that operates directly on noisy grayscale slices without explicit contour extraction, while the variational energy supplies explicit geometric preferences. Similarly, the neural representation contributes representation-dependent biases including a preference for lower-frequency structure, stochastic quadrature can under-sample narrow
    transition layers and thereby permit sharp interfaces, and the gradient-based optimization used to train the network couples the above effects and introduces a bias of its own. Across synthetic shapes, the Stanford Bunny, and scan-derived anatomical geometries, the resulting method produces smooth, connected reconstructions from sparse observations without offline training. Our experiments indicate that the interaction of explicit variational regularization with representational, sampling, and optimization biases can itself serve as a useful geometric reconstruction prior.
\end{abstract}



\section{Introduction}

\paragraph{} The need to reconstruct volumes from sparse data is shared by many fields. In the geosciences, the geometry of underground features can be estimated from surface measurements in order to understand a region's susceptibility to earthquakes \cite{famiani_geophysical_2020}. For geographers, it is necessary to reconstruct smooth interpolations of a landscape from the sparse contours of a topographic map \cite{hormann_c1-continuous_2003}. In computer graphics, the geometry of a 3D object must be inferred from a small number of photographs in order to render it from different perspectives \cite{mildenhall_nerf_2020}. Recently, volume reconstruction has even been explored as part of a system to guard 3D printers against cyberattacks \cite{mishra_real_2025}. Most common in the volume reconstruction literature, however, are applications involving biomedical imaging. Tomographic scans---typically from magnetic resonance imaging (MRI)---return planar grayscale images of biological structures which must be synthesized to reconstruct the 3D geometry. So-called "slice-to-volume reconstruction" (SVR) has been used to render 3D geometries of bones \cite{lin_measuring_2019}, organs \cite{young_fully_2024}, and tumors \cite{egger_gbm_2013}. With an eye to these biomedical applications, our focus will be on volume reconstruction applications where imaging data is given in slices and a 3D reconstruction of the volume of the imaged object is sought.

\paragraph{} SVR methods come in many forms, but can be broadly categorized by the way in which data is used. We refer to methods as "data-driven" which use data sets containing slice images as inputs and the corresponding 3D volume as the target to train machine learning models to do the reconstruction. These methods---which make use of sophisticated neural network architectures to build the map between slice data and the reconstructed volume---have been explored in the literature \cite{giannakopoulos_accelerated_2024, vishnevskiy_deep_2020, young_fully_2024}. An advantage of this approach is that the volume reconstruction process is extremely fast once the SVR model has been trained. A downside is that the reconstruction may be inaccurate if the slices involve features which were not seen in the training data. In this paper, we pursue an alternative: methods that reconstruct the volume directly from the given slice data of a specific case, without relying on prior training. These approaches treat volume reconstruction as an optimization problem tailored to the geometry at hand, offering greater flexibility when large annotated datasets are unavailable or when generalizing to novel geometric features is paramount.
 
\paragraph{} Interpolation-based methods are one such approach to volume reconstruction and have been researched for multiple decades. The earlier works build triangulations of the slice data, optimally connect nodes on adjacent slice surfaces to each other, and represent the surface of the volume as the set of piecewise planes which arise from the nodal connections \cite{hirsch_reconstruction_1997, boissonnat_three-dimensional_1993}. This represents one algorithmic approach to interpolating the surface from data specifying the boundary of the volume on slices. In order to find smoother surfaces, follow-up studies have used radial basis functions, B-splines, and other smooth bases to discretize a level set whose zero isocontour gives the boundary surface \cite{carr_reconstruction_2001, arigovindan_variational_2005, morse_interpolating_2001, cuomo_surface_2013, deng_surface_2012}. These methods are called interpolation-based because they use regression techniques to interpolate the boundary between slice planes. Because, by the definition of interpolation-based methods, there is no regularization applied to the regression problem, these methods struggle in SVR problems where slice data is widely spaced. One interpolation-based method which does apply some regularization to the inverse problem is presented in \cite{sawdayee_orex_2023}, where a neural network is used to represent a level set fit to slice data. However, as the authors' state, the neural network model class provides the prior on the reconstructed volumes, while the regularization is used primarily to aid in a post-processing step. In other words, there is no additional loss introduced to the optimization problem to enforce geometric properties of the boundary between measurement points or slice planes, so interpolations of widely-spaced measurements may give rise to boundaries with unphysical properties. Regardless of whether the surface position is stored explicitly (in the case of triangulation methods) or implicitly (in the case of level sets), the boundary of the volume is assumed to be known unambiguously in the slice planes. 

\paragraph{} Given that the objective of volume reconstruction is to determine the position of an unknown boundary, the assumption of labeled boundary points is often unrealistic. In general, there is no reason to assume that the imaging data provides measurements exactly on the boundary of the object whose volume is to be reconstructed. Given the presence of noise in the imaging process, we also cannot assume that the boundary is straightforward to identify even if it does coincide with measurement points. In fact, in the case of SVR from tomographic scans, the raw slice data comes in the form of noisy grayscale images which need to be preprocessed in order to estimate the position of the boundary. Image segmentation techniques must be used to find edges in the 2D slice data \cite{shu_variational_2025, spencer_variational_2016, cai_variational_2015, chen_introduction_2013}, but their output---which provides binary labels for the inside and outside phases of the image---may give the false impression that the boundary is sharp, when in fact there is ambiguity as to its position. As such, we wish to relax the assumption that any points on the boundary are known a priori. Furthermore, we seek an SVR method which works with the noisy grayscale images directly, and thus does not rely on image segmentation as a preprocessing step. To accomplish this, as well to regularize the regression problem to assist with widely-spaced slices, we turn to variational approaches to the volume reconstruction problem. These methods are often called "model-based," which distinguishes them from the data-driven and interpolation-based methods we have so far reviewed.

\paragraph{} By relaxing the assumption of the binary inside/outside labeling of the grayscale images, the phase of the material can be approximated by a continuous "phase field" variable $u(\mathbf{x}) \in [0,1]$, where $u=1$ represents the inside of the volume and $u=0$ represents the outside. Intermediate values of $u$ indicate transitions between the two phases. With the help of the phase field variable, variational volume reconstruction regularizes the interpolation of the slice data by minimizing an appropriately defined functional. This functional is used to simultaneously penalize intermediate values of the phase field and to enforce desired geometric properties of the reconstructed volume. While most authors in the literature agree on the double well potential term to penalize intermediate values of the phase field, there has been more variety in the approach to geometric regularization. Many works target the curvature of the reconstructed surface. In \cite{esedoglu_colliding_2013, bretin_volume_2017, kim_accurate_2015, droske_level_2004}, the "Willmore" energy is taken as the objective, which, by penalizing the norm of the gradient of the surface normal, gives rise to reconstructed surfaces of minimal mean curvature. These authors find a solution to the volume reconstruction problem by evolving the dynamical system corresponding to the gradient flow of the Willmore energy to steady state. Instead of targeting the curvature of the reconstructed surface, other works have used the surface area as an objective. The minimum surface area reconstruction is obtained with the "Cahn-Hilliard" energy \cite{chen_existence_1996}, and the volume reconstruction problem is solved by finding a minimum of this energy, whose gradient flow gives rise to the well-known "Allen-Cahn" equation \cite{li_three-dimensional_2015, bretin_penalized_2024, bretin_learning_2022, li_weighted_2022, kim_three-dimensional_2022}. Hybrid methods---which make use of the objectives for both the curvature and surface area---have also been proposed in the literature under the name of the "Euler-elastica" energy \cite{zhang_super-resolution_2024, zhu_image_2013}. We note that there have been other variational approaches which depart entirely from the curvature and surface area objectives given by the Willmore and Cahn-Hilliard energies \cite{heldmann_variational_2009, hong-kai_zhao_fast_2001}. 

\paragraph{} Our review suggests that the Cahn-Hilliard energy is the most popular objective for variational volume reconstruction from slices. It has been used successfully in the presence of more than two material phases \cite{li_multicomponent_2019}, to repair volumes with missing voxels \cite{li_efficient_2020}, and to continuously morph one shape into another \cite{han_simple_2023}. While the Cahn-Hilliard energy has been effective in many situations, it presents a number of issues that have yet to be addressed in the literature. The first problem we identify is the reliance on image segmentation to produce binary phase labels as a preprocessing step to variational volume reconstruction. We wish to devise a method which can reconstruct volumes directly from the grayscale images of slice planes, without arbitrary commitments as to the exact position of the boundary in noisy images prior to reconstructing the volume. This requires the ability to robustly interpolate intermediate phase values, where either measurement noise or inherent ambiguity of the boundary have introduced gray pixels into the slice images. Second, it is important for the SVR method to rely on as few slice planes as possible, as MRI data is time-consuming to obtain. This is another area where past methods struggle, as too few slice planes can lead to reconstructions with multiple disconnected geometries \cite{zhang_super-resolution_2024}. We note that there has been minimal exploration of reconstructing volumes from extremely sparse slice data. The third and final issue is that of computational cost. A drawback of model-based volume reconstruction is that the reconstruction is done on a case-by-case basis, which may lead to computational bottlenecks for analysts waiting on real-time results. Solving a nonlinear optimization problem for the phase field in three spatial dimensions is often expensive, and this cost exhibits problematic scaling as the mesh of a traditional finite element discretization is refined. 

\paragraph{} Reconstruction from sparse cross-sections is fundamentally ill-posed. Even when every observed
slice is matched exactly, infinitely many three-dimensional solids may interpolate the same data. The central question is therefore not whether a reconstruction method imposes a prior, but which prior it imposes. Classical variational methods make the prior explicit through a geometric energy and select a minimizer or stationary point of that model. Dataset-trained methods instead acquire a population-level prior from examples. In this work we investigate a training-data-free prior that emerges from the interaction of four ingredients: a chosen variational energy, a neural implicit representation, stochastic underintegration, and gradient-based optimization. These ingredients play distinct roles. The variational energy determines the geometric gradient information supplied to the optimizer; the neural representation changes which functional
structures are easy to represent and learn, including through spectral bias; stochastic underintegration changes the effective penalty experienced by narrow transition layers and can permit sharp interfaces; and the nonlinear optimizer couples these effects along a path through a nonconvex, stochastic objective. Our objective is therefore not to use a neural network as a high-fidelity numerical surrogate for a predetermined continuum minimizer. Rather, we exploit and study the combined algorithmic biases as a geometric reconstruction prior. With this in mind, our contributions in the work are as follows:

\begin{enumerate}
    \item We introduce a modified Cahn-Hilliard energy with anisotropic diffusion, a novel formulation of the measurement data penalty, and discretize the phase field with a neural network;
    \item We minimize the objective functional with ADAM optimization and, at each step, approximate integrals over the domain with Monte Carlo integration;
    \item We show that this method reconstructs complex volumes from noisy grayscale images without offline training data or image segmentation, effectively handles extremely sparse data, converges in seconds, and most importantly, leads to intuitive geometries that agree with the observed slice data while avoiding spurious high-curvature and disconnected features between slices.
\end{enumerate}

The rest of this paper is organized as follows. In Section 2, we introduce the problem setup, including a discussion of the noisy grayscale slice images we use in the SVR problem, a novel formulation of the data penalty, and the modified Cahn-Hilliard energy functional. In Section 3, we lay out the neural network discretization of the phase field and the Deep Ritz formulation of the SVR problem with random integration. In Section 4, we conduct a parameter study to provide guidelines for calibrating the model, compare our approach to three standard SVR methodologies, and showcase the method on a number of examples. Among these numerical examples are complex biomedical geometries, obtained from real 3D scans. In Section 5, we close with concluding remarks and directions for future work.


\section{Variational objective}

\subsection{Regression problem with noisy slice data}

\begin{figure}[hbt!]
\centering
\includegraphics[width=\textwidth]{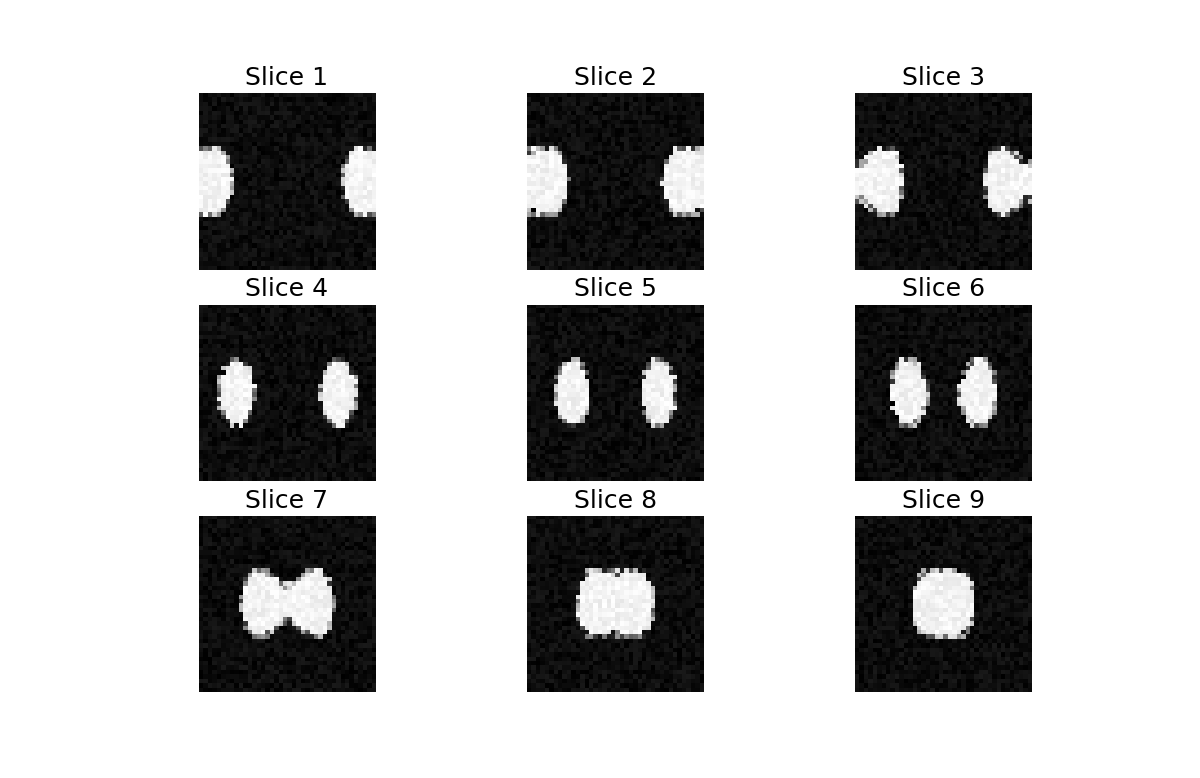}
\caption{Example slice data for a branching vessel geometry. The slice planes are sparse and the data is noisy. We relax the assumption that the precise location of a boundary curve is given within the slice planes.}
\label{slice_data}
\end{figure}

\paragraph{} First, we discuss the nature of the SVR measurement data and our formulation of the regression problem. Let $\mathbf{x} \in \Omega \subset\mathbb{R}^3$ be the spatial coordinate, $\Omega$ be the computational domain, and $u(\mathbf{x}) \in [0,1]$ be the phase field, where $u=1$ represents the inside of the volume and $u=0$ represents the outside. We define the computational domain to be the unit cube $\Omega=[0,1]^3$. We assume that imaging data of the volume of interest is provided in $S$ slice planes. We take the data from each slice plane to be a grayscale image in which the pixel values are scaled to lie in $[0,1]$ such that white corresponds to the inside phase $(u=1)$, black corresponds to the outside phase $(u=0)$, and gray pixels represent intermediate values. The gray pixels are a consequence of inherent ambiguity of the boundary position and noise in the imaging procedure. See Figure \ref{slice_data} for an example of such slice data. The objective functional for the variational volume reconstruction problem consists of two terms: (i) an error measure of the phase field with the measurement data and (ii) an energy-type functional which regularizes the regression problem by promoting certain geometric properties of the boundary surface. We now outline our formulation of the regression problem which is specifically designed for noisy grayscale images with ambiguous boundaries.

\paragraph{} Given that, in general, the grayscale images do not comprise binary pixel values, phase labels of the material are uncertain. We assume that the slice data is taken on a uniform rectangular grid in each slice plane. We define the set

\begin{equation*}
    \mathcal{S}_i = \{ \mathbf{x}^i_j\}_{j=1}^N,
\end{equation*}

\begin{figure}[hbt!]
\centering
\includegraphics[width=1.0\textwidth]{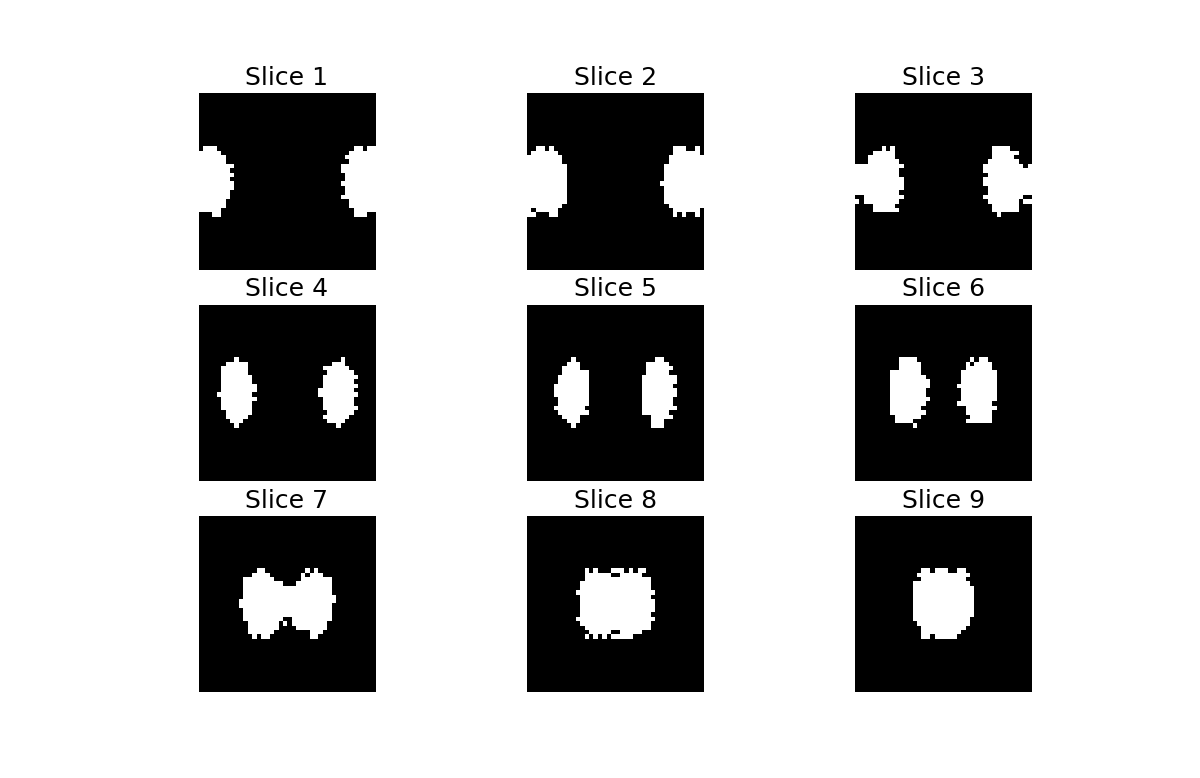}
\caption{Rounding the noisy slice data to binary phase labels leads to an arbitrary commitment to the boundary position and non-smooth boundary contours. }
\label{rounded_slice_data}
\end{figure}

\noindent where $\mathbf{x}_j^i$ represents the $j$-th point in the grid on the $i$-th plane and $N$ is the total number of points in the grid on each slice plane. The set of all measurement points is given by $\mathcal{S} = \{\mathcal{S}_i \}_{i=1}^S$ where $S$ is the number of slices. We must devise a regression problem governed by an objective of the following sort:

\begin{equation}\label{regress}
    \mathcal{R} = \frac{1}{2}\sum_{i=1}^S\Big(  \lVert u(\mathcal{S}^{\text{out}}_i) \rVert^2 + \lVert u(\mathcal{S}^{\text{in}}_i) - \mathbf{1}\rVert^2 \Big),
\end{equation}

\noindent where $\mathcal{S}_i^{\text{out}}$ and $\mathcal{S}_i^{\text{in}}$ respectively denote the sets of points determined to be outside and inside the volume in the $i$-th slice plane. How do we define these two sets from the noisy grayscale images without relying on image segmentation? As Figure \ref{rounded_slice_data} suggests, rounding the pixel values will give rise to a regression objective which enforces an artificially sharp boundary with a non-smooth contour. We propose an approach to remedy these issues in the following. First, we must define $\phi$ to be a scalar function which stores the imaging data. For example, $\phi(\mathcal{S}_i)$ provides the pixel values on the measurement grid for the $i$-th slice. We propose that the grayscale images are denoised through blurring, which mitigates the roughness of the boundary contour defined by rounding. Namely, we replace each pixel in each slice plane by the average of its neighbors. This is accomplished by running the following convolution kernel over the image:

\begin{equation*}
    \mathbf{K} = \frac{1}{9}\begin{bmatrix}
        1 & 1 & 1 \\ 1 & 1 & 1 \\ 1 & 1 & 1
    \end{bmatrix}.
\end{equation*}

\begin{figure}[hbt!]
\centering
\includegraphics[width=1.0\textwidth]{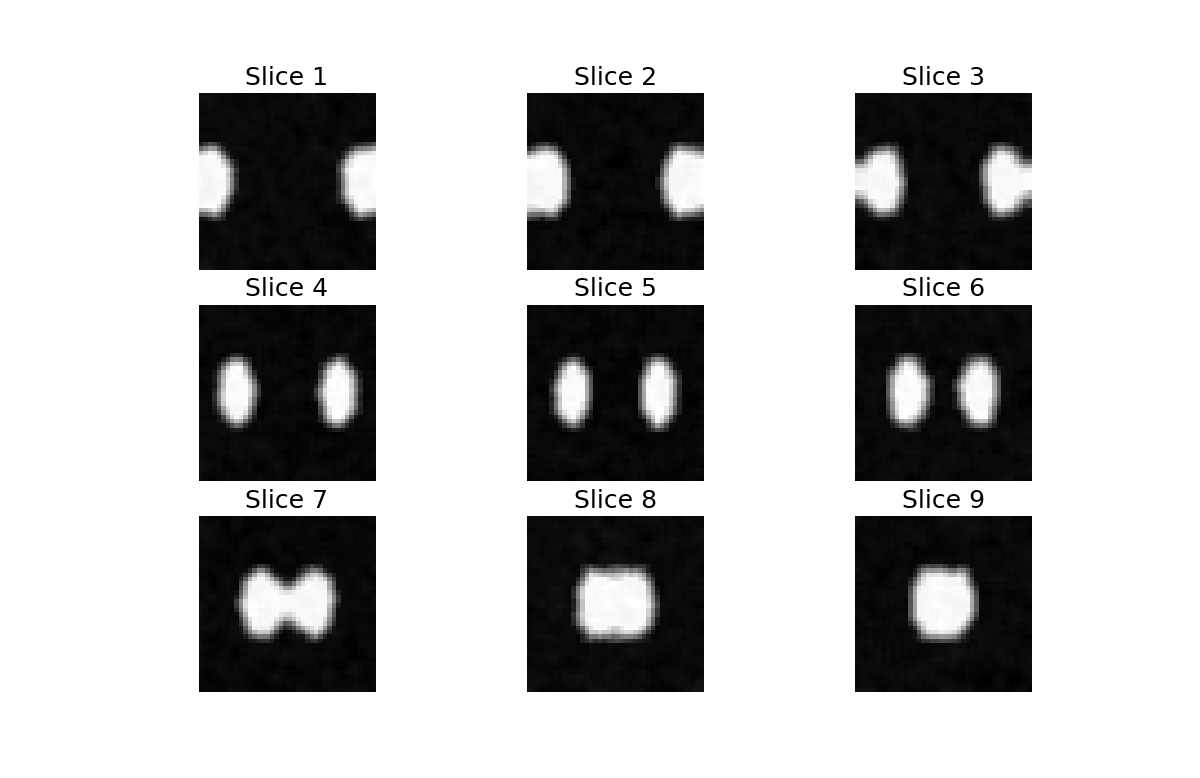}
\caption{Slice data from the branching vessel geometry blurred with an averaging convolution filter. This has the effect of denoising and smoothing the boundary contour defined by rounding.}
\label{blurred}
\end{figure}

See Figure \ref{blurred} for the blurred example slice data. We use symmetry boundary conditions on the convolution kernel to handle pixels at the edge of the image. The blurred image, which we denote $\hat \phi$, will have a smoother boundary contour under rounding, but there is still ambiguity as to the location of the boundary given the presence of gray pixels. Instead of creating a sharp boundary with a standard rounding procedure, we define the sets of outside and inside points with

\begin{equation}\label{sets}
    \mathcal{S}_i^{\text{out}} = \{ \mathcal{S}_i: \hat \phi(\mathcal{S}_i) < 1-c \}, \quad \mathcal{S}_i^{\text{in}} = \{ \mathcal{S}_i: \hat \phi(\mathcal{S}_i) \geq c \},
\end{equation}

\noindent where $c \in [0.5,1)$ is a threshold parameter which defines the tolerance for gray pixels in phase assignment. Note that Eq. \eqref{sets} operates on the blurred image and implies that $\mathcal{S}_i^{\text{out}} \cup \mathcal{S}_i^{\text{in}} \neq \mathcal{S}_i$, meaning that there are some points which are not assigned a phase. This definition respects our claim that the position of the boundary is not known a priori. The measurement points which are not assigned a phase do not show up in the regression objective of Eq. \eqref{regress}. By not assigning a phase to the intermediate pixel values, we rely on the geometric regularization to position the boundary of the volume.

\subsection{Regularization with modified Cahn-Hilliard energy}

\paragraph{} Minimizing Eq. \eqref{regress} with a particular discretization of the phase field yields an interpolation-based method for the volume reconstruction problem. When the slice data is sparse, additional regularization of the regression problem is required to obtain satisfactory properties of the boundary surface. In the literature, the Cahn-Hilliard energy is the most common strategy for this regularization. The standard Cahn-Hilliard energy functional reads

\begin{equation}\label{ch}
        \mathcal{E}'(u) = \int_{\Omega} \Bigg( \underbrace{\frac{\epsilon}{2} \lVert \nabla u \rVert^2 }_{\text{diffusion}} + \underbrace{\frac{1}{2\epsilon}u^2(1-u)^2}_{\text{double well potential}} \Bigg) d\Omega, 
\end{equation}

\begin{figure}[hbt!]
\centering
\includegraphics[width=0.99\textwidth]{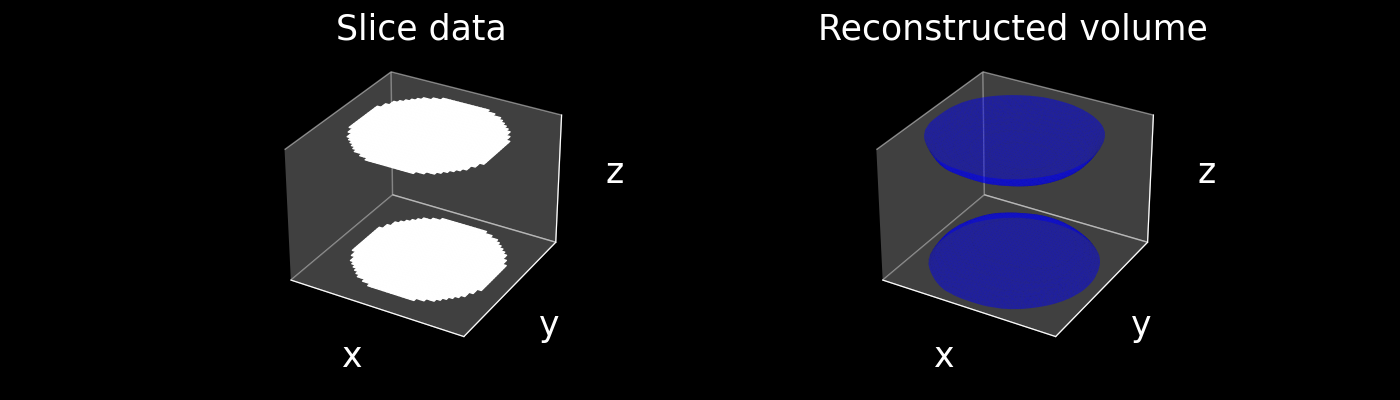}
\caption{With sparse slice data, the standard Cahn-Hilliard energy finds volumes of minimal surface area by reconstructing multiple disconnected volumes.}
\label{disconnected}
\end{figure}

\noindent where the diffusion term controls the width of interface of the volume by penalizing gradients, the double well potential penalizes intermediate values of the phase field, and the length scale parameter $\epsilon$ controls their respective weighting. It is known that as a regularizer to a regression problem, the minimum of Eq. \eqref{ch} corresponds to volumes of minimum surface area \cite{chen_existence_1996}. This means that when slice data is widely spaced, Eq. \eqref{ch} finds minimal surfaces by reconstructing disconnected volumes. This issue is noted by the authors in \cite{zhang_super-resolution_2024} and can be seen in Figure \ref{disconnected}. The formation of two disconnected regions is a valid mathematical solution to the volume reconstruction problem, but often not what the analyst desires. As such, we seek an objective which can be tuned to obtain the disconnected regions of Figure \ref{disconnected} at one setting, but will reconstruct a circular cylinder at another setting. In other words, we must modify the Cahn-Hilliard energy in such a way that the formation of disconnected regions can be discouraged. This is accomplished by introducing anisotropic diffusion, which penalizes gradients of the phase field in the direction normal to slice planes more severely. The modified Cahn-Hilliard energy reads

\begin{equation}\label{CHmod}
\begin{aligned}
    \mathcal{E}(u) = \int_{\Omega} \frac{1}{2}  \nabla u^T \cdot \boldsymbol \epsilon \nabla u  + \frac{1}{2\bar \epsilon}u^2(1-u)^2 d\Omega,\\
    \boldsymbol \epsilon = \text{diag}(\epsilon_x,\epsilon_y ,\epsilon_z ),\\
    \bar \epsilon = \text{tr}(\boldsymbol \epsilon)/3.
\end{aligned}
\end{equation}

Going forward, we assume that slice data lies in the $x$-$y$ plane and that $\epsilon_x=\epsilon_y$, as the measurements are equally spaced within the slice planes. When the slice planes are sparse, we expect that $\epsilon_z>\epsilon_x,\epsilon_y$ will discourage the formation of disconnected regions by increasing the penalty on $|\partial u / \partial z|^2$. This anisotropy biases the boundary surface away from minimal surface area and toward being vertical between slices. The choice of values for the free parameters $[\epsilon_x,\epsilon_y,\epsilon_z]$ will be discussed in Section 4. Given that it scales spatial gradients, we emphasize that our forthcoming guidance on the choice of $\boldsymbol \epsilon$ will be specific to $\Omega = [0,1]^3$. This is without any loss of generality, as slice data can be rescaled to lie in the unit cube. 

\subsection{Variational volume reconstruction with noisy, sparse slice data}

\paragraph{} With phase labels assigned to measurement points per Eq. \eqref{sets}, we combine Eqs. \eqref{CHmod} and \eqref{regress} to form the objective functional for the volume reconstruction problem:

\begin{equation}\label{obj}
\begin{aligned}
    \Pi(u) = \frac{p}{S^*}\mathcal{R}(u) + \mathcal{E}(u)
     = \frac{p}{2S^*}\sum_{i=1}^{S}\Big(  \lVert u(\mathcal{S}^{\text{out}}_i) \rVert^2 + \lVert u(\mathcal{S}^{\text{in}}_i) - \mathbf{1}\rVert^2 \Big) 
     \\+    \int_{\Omega} \frac{1}{2}  \nabla u^T \cdot \boldsymbol \epsilon \nabla u  + \frac{1}{2\bar \epsilon}u^2(1-u)^2 d\Omega,
\end{aligned}
\end{equation}

\noindent where $S^*=|\mathcal{S}^{\text{out}} \cup \mathcal{S}^{\text{in}}|$ is the number of points with assigned phase and $p$ is a penalty parameter that controls the weighting between the regression problem and the regularization with the modified Cahn-Hilliard energy. In addition to the anisotropic diffusion, we give guidelines for the choice of the penalty parameter in Section 4. Eq. \eqref{obj} is designed to handle noisy and sparse slice data with no prior knowledge of the boundary position. A solution to the volume reconstruction problem is obtained by minimizing this objective functional:

\begin{equation}\label{solution}
    u^*(\mathbf{x}) = \underset{u(\mathbf{x})}{\text{argmin }} \Pi(u).
\end{equation}

Once the solution is obtained, we take the reconstructed volume $\mathcal{V}$ and its boundary $\partial \mathcal{V}$ to be given by

\begin{equation}\label{geos}
    \mathcal{V} = \Big \{ \mathbf{x} : u^*(\mathbf{x}) \geq 0.5 \Big \}, \quad \partial \mathcal{V} = \Big \{ \mathbf{x} : u^*(\mathbf{x}) = 0.5 \Big \}.
\end{equation}

Having formulated the continuous volume reconstruction problem, we now must discretize the phase field in order to obtain numerical solutions. Prior work in the variational volume reconstruction literature has generally relied on grid- or mesh-based discretizations of the phase field. Thus, in addition to the noise-informed penalty and anisotropic diffusion in Eq. \eqref{obj}, another point of departure of our work is to discretize the phase field with a neural network. One exception to this trend is \cite{walker_crosssdf_2025}, where the authors discretize the phase field with a neural network, though they use a minimum surface formulation analogous to the standard Cahn-Hilliard objective functional. In the next section, we lay out the neural network discretization and optimization strategy before showcasing the benefits of our proposed method. 


\section{Deep Ritz method with random integration}

\paragraph{} The variational volume reconstruction problem given by Eqs. \eqref{obj} and \eqref{solution} aligns with many of the strengths of neural network discretizations of fields. This problem has the following features: 1) the geometry of the computational domain $\Omega$ is rectangular, 2) there are no boundary conditions, 3) the solution is naturally given by the minimizer of an objective, 4) the optimization problem is nonlinear, 5) the solution exhibits sharp gradients at locations which are not known a priori, and 6) the solution field is constrained to lie in a certain range (i.e., $u(\mathbf{x}) \in [0,1]$). Features 1-2 are assets because, without a mesh, handling complex geometries and the associated boundary conditions for neural network discretizations is challenging \cite{sukumar_exact_2022}. In SVR, the volume is embedded in a rectangular domain and the phase field has no boundary conditions. Feature 3 is beneficial because, as is well known, neural networks are trained by solving optimization problems. Feature 4 states that finding a stationary point of the objective function requires a nonlinear solve even with traditional spectral and finite element discretizations, meaning that the neural network is not the sole source of nonlinearity. This contrasts with many neural network-based solutions to PDEs, where the optimization problem becomes nonlinear solely from the choice of discretization \cite{liu_deep_2023}. Feature 5 is handled well by neural networks because of their natural "adaptivity," meaning that the neural network can resolve high-gradient regions of the solution without prior knowledge of their location \cite{manav_phase-field_2024}. Finally, we will see that the constraint implied by feature 6 can be built into the neural network architecture and thereby satisfied automatically.

\paragraph{} Using neural networks to discretize fields that minimize variational objectives was first explored with the "Deep Ritz method" (DRM) \cite{e_deep_2017}. With DRM, the neural network is used to discretize the unknown solution to a PDE with an associated variational principle. DRM has been studied extensively in the context of engineering mechanics, for example in linear elasticity \cite{linghu_higher-order_2023}, hyperelasticity \cite{abueidda_deep_2022}, and fracture mechanics \cite{manav_phase-field_2024}. Because the Cahn-Hilliard energy in Eq. \eqref{CHmod} is a variational objective much like the energy functionals of mechanics, we take our neural network discretization of the phase field used in the SVR problem to be an application of the DRM.

\paragraph{} We use a multilayer perceptron (MLP) neural network to discretize the phase field in Eq. \eqref{obj}. With this architecture, the input-output relation for the $i$-th hidden layer of the network is given by

\begin{equation*}
    \mathbf{y}_i = \sigma_i\Big(  \mathbf{W}_i\mathbf{y}_{i-1} + \mathbf{b}_i  \Big), \quad i=1,2,\dots,L,
\end{equation*}

\noindent where $L$ is the number of layers, $\mathbf{y}_i$ represents the value of the neurons in the $i$-th layer, and $\sigma_i(\cdot)$ is a nonlinear "activation function" which is applied element-wise to the neurons of layer $i$. As shown, the output $\mathbf{y}_i$ then becomes the next layer's input. The input layer is taken to be the spatial coordinate, i.e. $\mathbf{y}_0 = \mathbf{x}$, and the scalar output layer is the solution field $y_L = u(\mathbf{x})$. The parameters of the neural network are the collection of the weight matrices $\mathbf{W}_i$ and bias vectors $\mathbf{b}_i$ for each layer, with the collection of all parameters written as $\boldsymbol{\theta} = [ \mathbf{W}_1, \mathbf{b}_1, \mathbf{W}_2,\mathbf{b}_2,\dots]$. Independent of the activation functions at the previous layers of the network, we take the activation function at the final layer to be

\begin{equation*}
    \sigma_L(\cdot) = \frac{1}{2}\Big(  \tanh(\cdot) + 1\Big),
\end{equation*}

\noindent which ensures that $u(\mathbf{x}) \in [0,1]$ by construction. Denoting the discretized phase field as $\hat u(\mathbf{x};\boldsymbol \theta)$, the variational objective becomes

\begin{equation}\label{objhat}
\begin{aligned}
    \hat \Pi(\boldsymbol \theta) = \frac{p}{2S^*}\sum_{i=1}^S\Big(  \lVert \hat u(\mathcal{S}^{\text{out}}_i) \rVert^2 
    \\+ \lVert \hat u(\mathcal{S}^{\text{in}}_i) - \mathbf{1}\rVert^2 \Big) +    \int_{\Omega} \frac{1}{2}  \nabla \hat u^T \cdot \boldsymbol \epsilon \nabla \hat u  + \frac{1}{2\bar \epsilon}\hat u^2(1-\hat u)^2 d\Omega.
\end{aligned}
\end{equation}

The numerical solution $ \hat u^*(\mathbf{x})$ to the SVR problem is obtained by finding parameters $\boldsymbol \theta^*$ such that Eq. \eqref{objhat} is minimized. As is common for training neural networks, we use ADAM optimization to find a minimum of the objective function. One potential disadvantage of DRM is that it may take thousands of optimization steps (epochs) to converge \cite{e_deep_2017}. Already in three spatial dimensions, the number of integration points to accurately compute the integral in Eq. \eqref{objhat} is large. This issue is especially pronounced because we cannot rely on quadrature rules to efficiently integrate neural networks. As such, a standard choice of integration grid would be a tensor product of uniformly spaced integration grids. Yet even with a modest $50$ integration points per coordinate direction, the integration grid requires $1.25\times 10^5$ evaluations of the neural network to compute the objective at each optimization step. Given the high computational cost of evaluating these integrals, and our expectation that the optimizer requires between $10^3$ and $10^5$ steps to converge, this method will be significantly more expensive than finite element implementations.

\paragraph{} In order for a volume reconstruction method to be feasible for use in real time, the reconstruction process must take seconds to minutes, at most. Thus, low computational cost is an important component of a successful SVR method. To remedy the prohibitively large number of function evaluations to find a minimum of Eq. \eqref{objhatmc} with full integration and ADAM optimization, we propose the following: instead of fully integrating the objective at each optimization step, we apply Monte Carlo integration using randomly sampled batches, analogous to minibatching the training data in stochastic gradient descent. At each optimization step, the objective is approximated as

\begin{equation}\label{objhatmc}
\begin{aligned}
    \hat \Pi(\boldsymbol \theta) \approx \frac{p}{2S^*}\sum_{i=1}^S\Big(  \lVert \hat u(\mathcal{S}^{\text{out}}_i) \rVert^2 + \lVert \hat u(\mathcal{S}^{\text{in}}_i) - \mathbf{1}\rVert^2 \Big) 
    \\+    \frac{1}{B} \sum_{i=1}^B \Big( \frac{1}{2}  \nabla \hat u(\mathbf{x}_i)^T \cdot \boldsymbol \epsilon \nabla \hat u(\mathbf{x}_i)  + \frac{1}{2\bar \epsilon}\hat u(\mathbf{x}_i)^2(1-\hat u(\mathbf{x}_i))^2 \Big),
\end{aligned}
\end{equation}

\noindent where $B$ is the batch size and $\mathbf{x}_i \overset{\text{i.i.d}}{\sim} \mathcal{U}(\Omega)$. Note that the regression loss, which is evaluated only at the measurement points, is not approximated via batching and is thus computed exactly at each step. The volume reconstruction problem optimization problem is thus

\begin{equation}\label{stochastic_drm}
    \underset{\boldsymbol \theta}{\text{argmin }} \hat \Pi(\boldsymbol \theta),
\end{equation}

\noindent where gradients of Eq. \eqref{objhatmc} are computed using automatic differentiation with PyTorch. While Monte Carlo integration is motivated first by computational cost, we will show that the inaccuracies of the integral computation actually improve the quality of the reconstruction. Of course, this relies on the assumption that the variational energy is only a means to the end of regularizing the regression to slice data---our concerns lie in obtaining high-quality reconstructions, rather than obtaining the continuum minimum energy solution. As there is neither a mesh nor a fixed integration grid, this method is simple to implement. In order for this method to be effective, we must demonstrate that any increase in the number of steps to convergence as a result of error in gradient computations does not overshadow the reduction in function evaluations per step. In the following section, we demonstrate the effectiveness of the stochastic objective given by Eq. \eqref{objhatmc} on a number of examples. These examples show that convergence occurs at a satisfactory rate despite imprecise gradient computations.

\paragraph{} We emphasize that the solution to Eq. \eqref{stochastic_drm} obtained with numerical optimization need not match the solution to the variational problem of Eq. \eqref{solution}. This phenomenon is observed in subsequent numerical examples. Our experience suggests that minima of Eq. \eqref{stochastic_drm} can be understood through the interaction of the Deep Ritz regularization, the stochastic integration, and the inductive biases of neural networks. The double well term in the Deep Ritz regularization encourages binary phases, thus, the network can drive the objective down by choosing the phase labels according to the penalty term. There is no inherent conflict between the penalty and the double well terms in the objective. However, the diffusion term penalizes rapid transitions from one phase to another with the magnitude of the gradient of the phase field. There are potential trade-offs between minimizing phase field gradients and enforcing binary phases in accord with the data. But, recalling that we randomly sample integration points, the probability of an integration point landing in the transition region decreases with the width of this region. This creates an optimization incentive for transition regions to become sufficiently narrow that they are infrequently sampled, reducing the effective diffusion penalty seen by the stochastic objective. In our method, the potential for the contribution of the boundary to vanish in forming the objective represents a significant departure from the solution of the continuum variational problem. An additional ingredient in this method is the so-called ``spectral bias'' of neural networks, which is typically understood as the tendency of networks to fit functions by progressing from low- to high-frequency Fourier components \cite{rahaman_spectral_2019, xu_frequency_2020}. In our case, the spectral bias provides one plausible explanation for the comparatively simple boundary geometry obtained between slices. The preference of networks to fit these parsimonious, low-frequency functional forms has been discussed in the context of decision boundaries for classification problems \cite{huang_understanding_2020}. In this work, it is noted that complex decision boundaries correspond to solutions with infinitesimally small volume in the loss landscape, meaning the optimizer does not find them in practice. Given that the volume reconstruction problem is closely connected to a two-class classification problem, we believe the insights of these authors regarding the inductive biases of neural network to be applicable here. In this sense, the stochastic Deep Ritz algorithm should be viewed as a method to reconstruct geometries from sparse slices, rather than a faithful numerical approximation of the continuum variational solution. 

\paragraph{} Because slice-to-volume reconstruction from sparse slices is fundamentally ill-posed, our goal is not exact recovery of a unique ground-truth volume from undersampled data. Rather, we seek reconstructions that agree with the available slice measurements while exhibiting desirable geometric behavior between slices, such as smoothness, connectivity, and the avoidance of spurious high-curvature or disconnected features. Along these lines, that our method is not expected to obtain the continuum variational solution to Eq. \eqref{solution} is no cause for concern. Monte Carlo integration allows the network to learn a transition boundary with small volume so that the diffusion term does not appear in Eq. \eqref{stochastic_drm}. Thus, the phase field is constant in the regions suggested by the penalty term with a sharp boundary separating these regions, where the shape of the boundary is expected to be parsimonious according to the spectral bias and the preference of the optimizer for solutions with large volume in the loss landscape. We note that the variational objective is meant only to regularize the regression to the slice data, meaning that we see no reason to prefer the continuum variational solution over those obtained by our method. We remind the reader that we treat the entire algorithm as a prior on reconstructed geometries. The variational energy supplies explicit geometric preferences; the network changes which functional structures are easy to represent and learn; finite-batch quadrature changes the effective penalty experienced by narrow transition layers; and the optimizer couples these effects along a path through a stochastic, nonconvex objective. We
therefore treat the complete algorithm---rather than the continuum functional alone---as the
reconstruction model. As the following numerical examples show, this method consistently leads to intuitive, smooth, and connected reconstructed volumes.


\section{Numerical examples}

\subsection{Parameter study}

\begin{figure}[hbt!]
\centering
\includegraphics[width=0.5\textwidth]{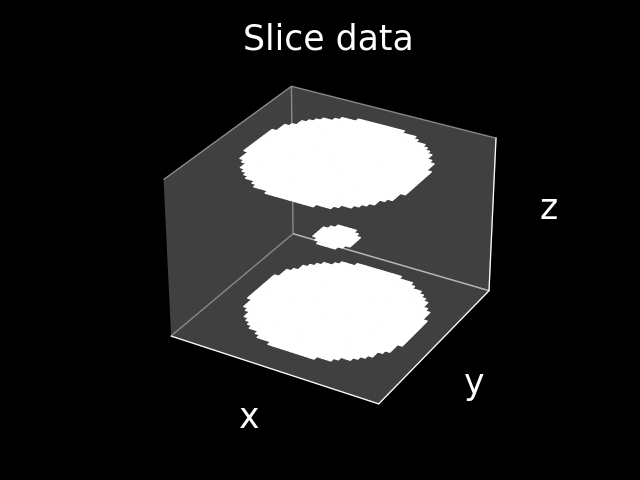}
\caption{Hourglass geometry used to conduct a parameter study to tune our proposed volume reconstruction method. We study the effects of the components of the diffusion tensor $\boldsymbol \epsilon$, the penalty $p$, the integration batch size $B$, and the number of epochs of ADAM optimization on qualitative features of the reconstructed geometries.}
\label{hourglass}
\end{figure}

\paragraph{} To effectively deploy our model, we must first set guidelines for the choice of the in-plane diffusion coefficients $\epsilon_x=\epsilon_y$, the out-of-plane diffusion $\epsilon_z$, the penalty parameter $p$, the Monte Carlo integration batch size $B$, and the number of optimization epochs. Noting that the entries of the diffusion tensor scale spatial gradients of the phase field, we reiterate that our choice of $[\epsilon_x,\epsilon_y,\epsilon_z]$ assumes that $\Omega=[0,1]^3$. In order to provide guidelines for parameter selection, we conduct a study reconstructing an hourglass geometry from equally spaced slice planes. In particular, we take $S=3$ slices and $N=1600$ noiseless measurements of the phase in each slice plane. The level set used to generate the hourglass geometry is 

\begin{equation*}\label{hourglassls}
    \Phi(x,y,z)=(x-0.5)^2 + (y-0.5)^2 - 3(z-0.5)^4/2-0.01.
\end{equation*}

See Figure \ref{hourglass} for the sparse slice data from this level set. We claim that there is no single quantitative metric that adequately captures the success or failure of the reconstructed volumes. For example, it is not useful to compute an error of the reconstructed boundary surface with the zero isocontour of the level set, as there is no reason to expect exact recovery of the true surface from sparse slice data. Instead, we will sweep over parameter settings and qualitatively assess the boundaries of the reconstructed volumes. In particular, we are interested in whether multiple disconnected regions form and the geometric features of the boundary surface(s).

\begin{table}[hbt!]
\centering
\caption{The 9 settings used in our parameter study. The parameter setting number is used to index the results shown in Figures \ref{sweep} and \ref{losses}.}
\label{tab:sweep}
\begin{tabular}{cccccccl}
    \toprule
     \textbf{Parameter setting} & $\epsilon_x=\epsilon_y$ & $\epsilon_z$ & $p$ & $B$ & epochs  \\ 
     \midrule
    1 & 1 & 1 & 10 & 5000 & 5000 \\ 
    2 & 1 & 0.1 & 1000 & 5000 & 5000 \\ 
    3 & 1 & 100 & 1000 & 5000 & 5000 \\ 
    4 & 1 & 5 & 1000 & 5000 & 5000 \\ 
    5 & 1 & 5 & 2500 & 5000 & 5000 \\ 
    6 & 1 & 5 & 1000 & 10000 & 5000 \\ 
    7 & 1 & 5 & 1000 & 5000 & 10000 \\ 
    8 & 1 & 2.5 & 2500 & 2500 & 10000 \\ 
    9 & 1 & 2.5 & 2500 & 10000 & 2500 \\
    \bottomrule
\end{tabular}
\end{table}

\begin{figure}[hbt!]
\centering
\includegraphics[width=1.0\textwidth]{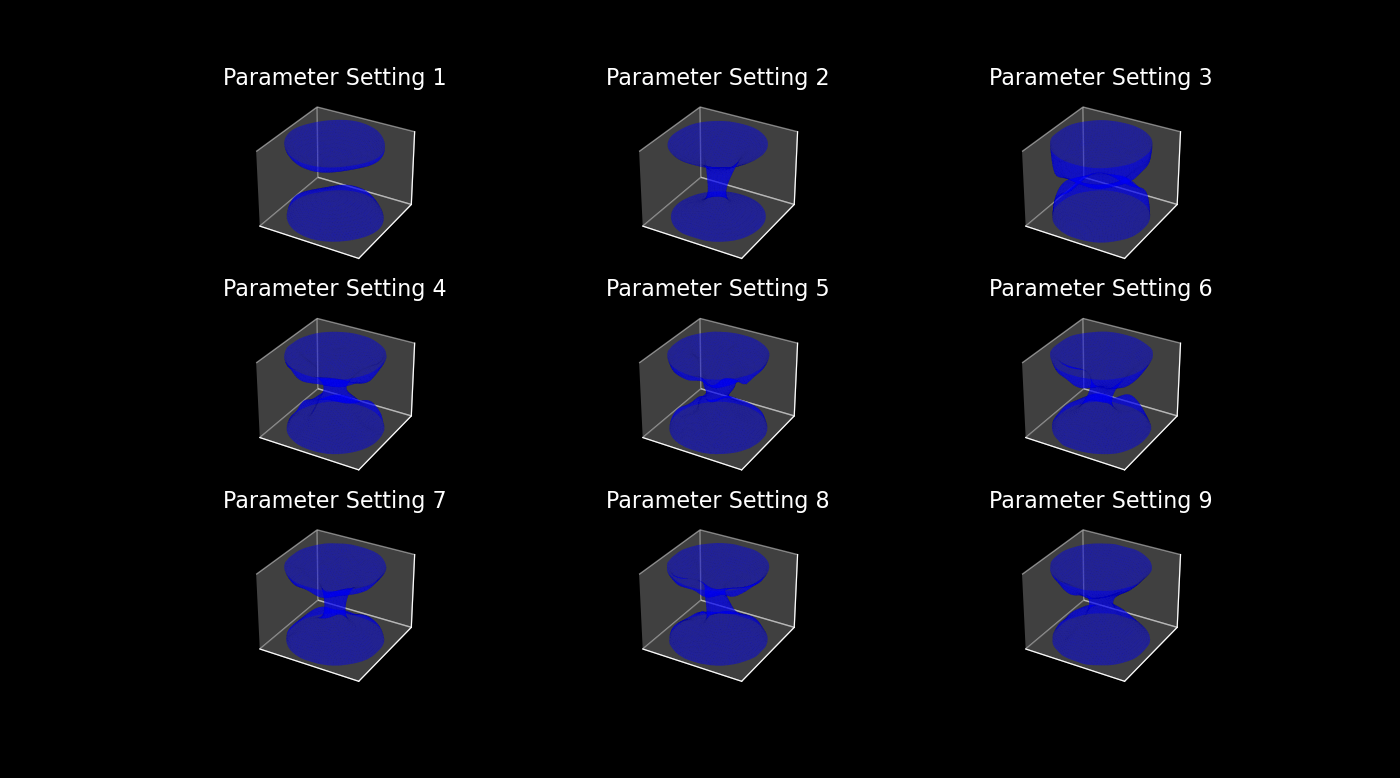}
\caption{The $\hat u^*(\mathbf{x})=0.5$ isosurface that defines the boundary of the reconstructed volume at the 9 parameter settings. Significant qualitative differences in the reconstructed geometry are observed as a result of the choice of penalty parameter $p$ and out-of-plane diffusion $\epsilon_z$. }
\label{sweep}
\end{figure}

\begin{figure}[hbt!]
\centering
\includegraphics[width=1.0\textwidth]{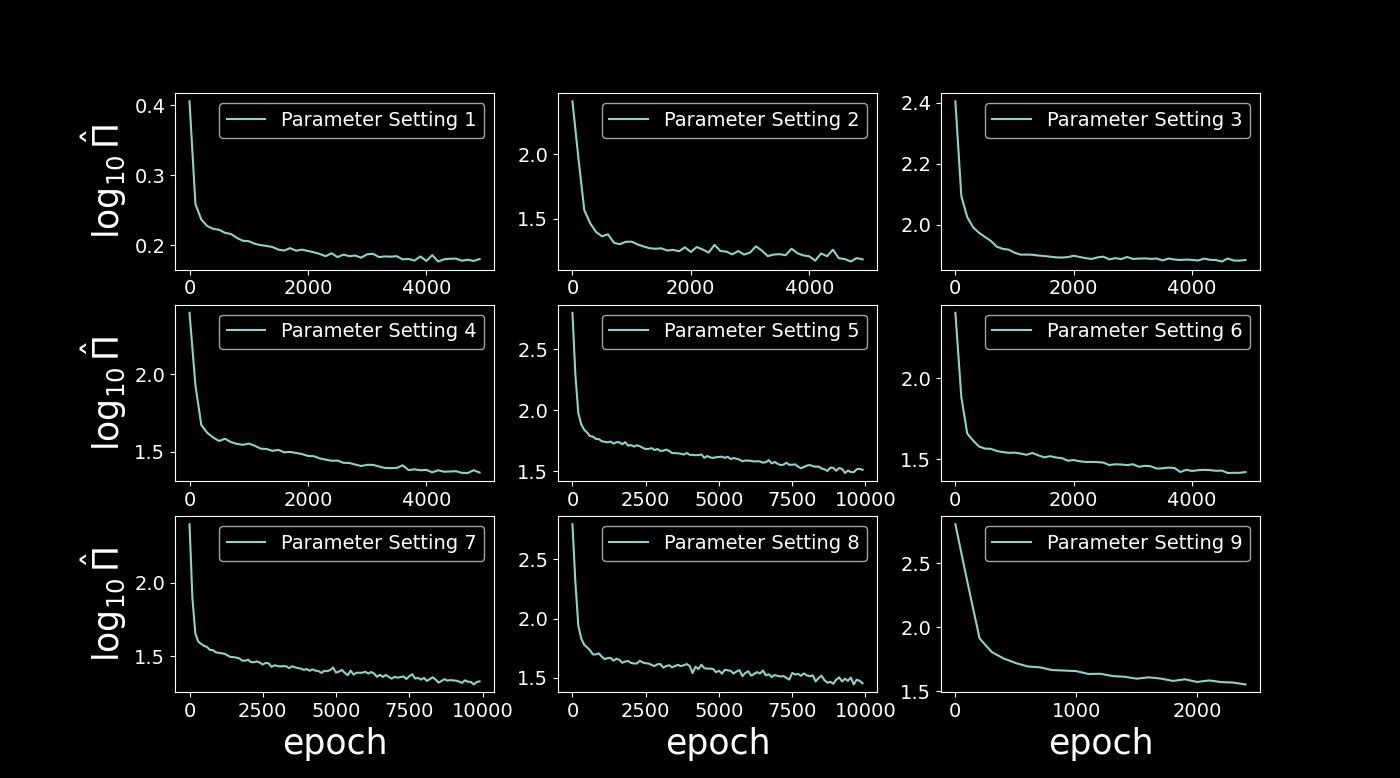}
\caption{Convergence of the stochastic objective used to reconstruct the hourglass volume from three slices. The isosurface plots suggest that the slight decreases in the objective in the 5000-10000 epoch range do not influence the qualitative features of the reconstructed volume.}
\label{losses}
\end{figure}

\paragraph{} To avoid a prohibitively high-dimensional search space, we decide to fix $\epsilon_x=\epsilon_y=1$ and sweep over the remaining parameters. Note that in many phase field models, parameters such as the components of $\boldsymbol \epsilon$ are length scales which define the width of the transition from one phase to another \cite{ghaffari_motlagh_deep_2023}. Thus, in order to control the geometry of the transition region in and out of slice plane, we might think to define the components of $\boldsymbol \epsilon$ in terms of the distance between measurement points in plane and between the slices. However, numerical experimentation indicated that suitable problem parameters could be determined without non-dimensionalizing in this way. See Table \ref{tab:sweep} for the parameter settings used in our study. The results of the experiments are reported with the boundaries of the reconstructed volumes in Figure \ref{sweep}, and the training convergence in Figure \ref{losses}. With parameter setting 1, the penalty on the measurement data is not sufficiently large to prevent the formation of two disconnected regions, as the minimizer of the variational objective does not enforce agreement with the small set of measured interior points on the $z=0.5$ slice. Parameter setting 2 has a large penalty on the data but a small penalty on out-of-plane diffusion, which gives rise to a minimal surface solution of three disconnected regions. Parameter setting 3 shows that too large a penalty on the out-of-plane diffusion prevents $z$ gradients of the boundary from forming, leading to a geometry resembling two stacked cylinders. Parameter settings 4-9 explore smaller variations to the out-of-plane diffusion, the data penalty, the batch size, and the number of epochs. All of these geometries are hourglass shaped, thus qualitatively reproducing the original volume. With these results in mind, this brief study motivates the following heuristics to guide the selection of parameters: (i) the penalty $p$ must be large $(p \geq 1000)$ in order to enforce agreement with the data, (ii) the out-of-plane diffusion should be set as 1-10$\times$ the in-plane diffusion when the slices are sparse, (iii) obtaining exact convergence of the minimization problem is not essential for satisfactory reconstructions (see Figure \ref{losses}), and (iv) the way in which function evaluations are distributed between the integration batch and number of epochs is less important than the total number of evaluations. See Table \ref{tab:recommended} for a summary of the recommended parameter settings. The following numerical examples are used to demonstrate the effectiveness of these guidelines.

\begin{table}[hbt!]
\centering
\caption{Recommended parameter settings based on the parameter study of the hourglass geometry.}
\label{tab:recommended}  
\begin{tabular}{ccl}
    \hline
    \textbf{Parameter} & \textbf{Recommended setting} \\
    \hline
    $\epsilon_x = \epsilon_y$ & 1 \\
    $p$ & $\geq 1000$ \\
    $\epsilon_z$ & $2.5$–$10$ \\
    $B$ & 5000–10000 \\
    epochs & 5000–10000 \\
    \hline
\end{tabular}
\end{table}


\subsection{Cylinder from two slices}

\paragraph{} We now apply the parameter recommendations from the previous study to a new test case. In particular, we explore the reconstruction of a cylindrical volume from only two slices. The level set used to generate the cylinder is

\begin{equation*}
    \Phi(x,y,z) = (x-0.5)^2+(y-0.5)^2-0.2.
\end{equation*}

\begin{figure}[hbt!]
\centering
\includegraphics[width=0.99\textwidth]{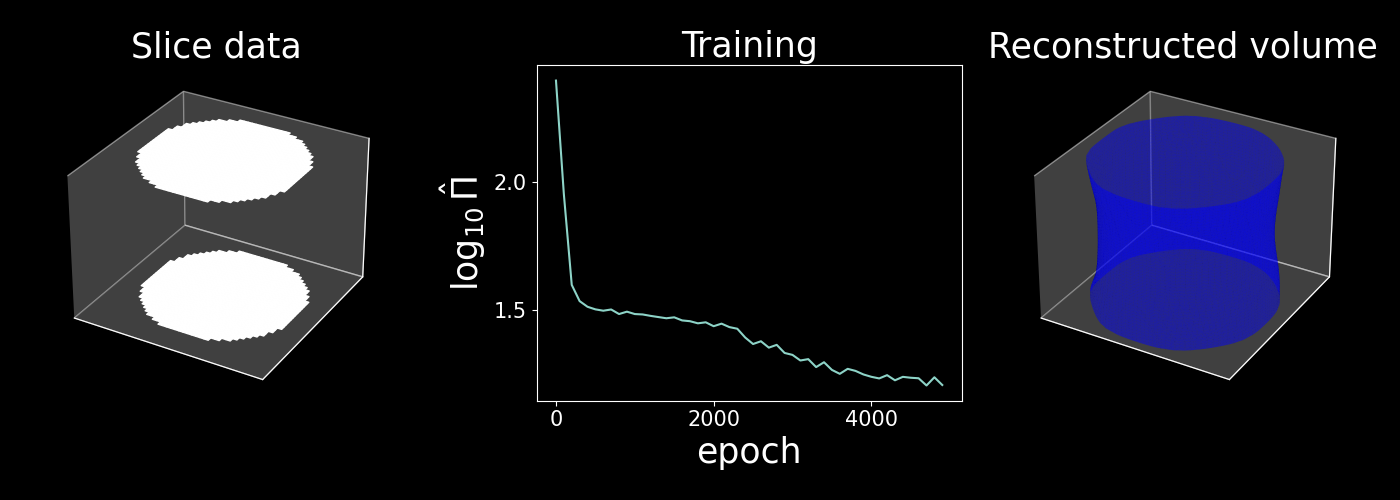}
\caption{Reconstructing a cylinder from two noiseless slice planes. The anisotropic diffusion penalty discourages the minimal surface solution of two disconnected regions.}
\label{cylinder}
\end{figure}

The two slice planes from which we obtain phase data are located at $z=0$ and $z=1$. We again work with $N=1600$ noiseless measurement points per slice plane. Given the results of the parameter study, we choose $p=1000$ and $\epsilon_z=10$ due to the wide spacing of the slices. We take the integration batch size to be $B=5000$ and run ADAM for $5000$ epochs. The results of the volume reconstruction are shown in Figure \ref{cylinder}. Despite slight hourglass-shaped narrowing of the geometry in the center of the domain, the anisotropic diffusion penalty discourages the formation of two disconnected regions and leads to satisfactory reconstruction of the cylinder. For reference, Figure \ref{disconnected} shows the minimal surface solution with two disconnected regions obtained from isotropic diffusion ($\epsilon_z=1$). This problem has a run time of 28.8 seconds, which we consider suitable for clinical use. We note that further increases in performance could be obtained with parallelization of the training.  

\begin{figure}[hbt!]
\centering
\includegraphics[width=0.99\textwidth]{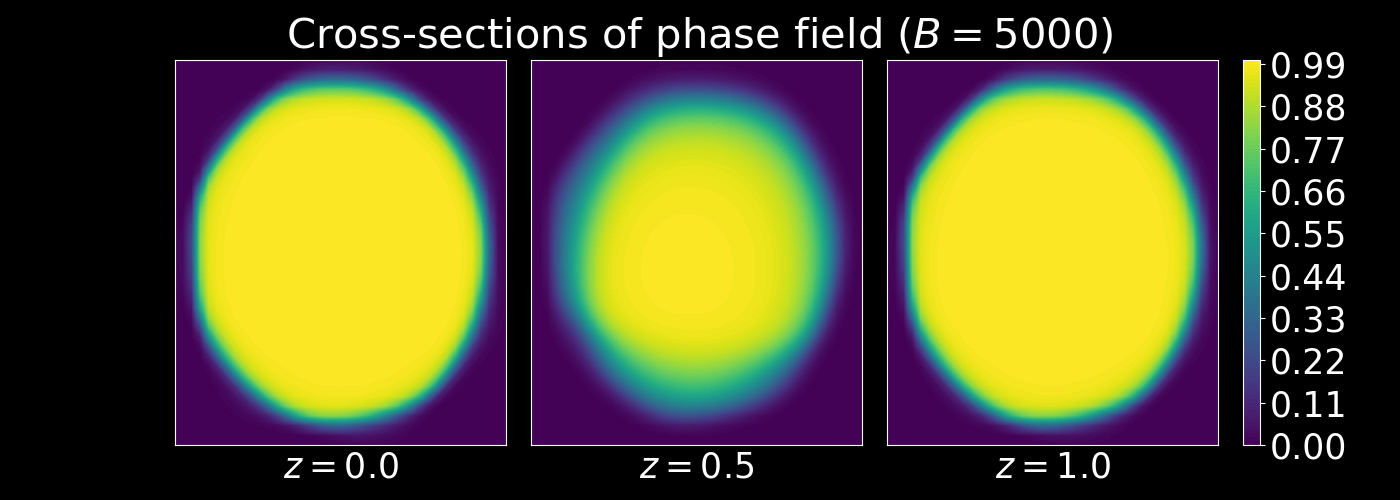}
\caption{The two slice planes with measurement data have crisp boundaries as a result of the large penalty parameter, but the boundary of the interior slice is diffuse.}
\label{contour}
\end{figure}

\begin{figure}[hbt!]
\centering
\includegraphics[width=0.99\textwidth]{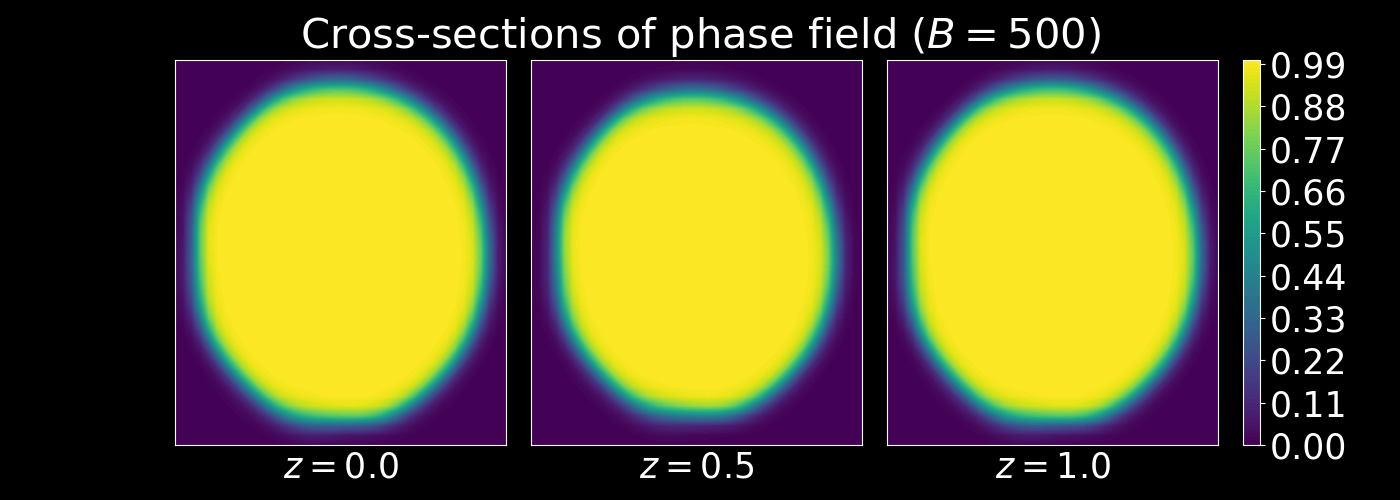}
\caption{Sharper transitions between the inside and outside phase are observed in the $z=0.5$ cross-section when the integration batch is reduced to $B=500$.}
\label{small_batch}
\end{figure}

\paragraph{} Figure \ref{cylinder} shows the $\hat u^*(\mathbf{x})=0.5$ isosurface as the boundary of the reconstructed geometry, but such plots do not give insight into the width of the transition region between the inside and outside phase of the material. See Figure \ref{contour} for heat maps of cross-sections of the converged phase field. The width of the transition region is significantly larger at the $z=0.5$ cross-section than the $z=0$ and $z=1$ slice planes, on which the regression loss enforces a sharp transition. The sharpness of the transition region in cross-sections which do not contain measurements is one way to explore the accuracy of the stochastic approximation of the variational objective given by Eq. \eqref{objhatmc}. To see this, consider the following: as the transition becomes sharper, the transition region occupies an increasingly small fraction of the volume of $\Omega$. Thus, it becomes increasingly unlikely that the Monte Carlo integration points will fall in the transition region. If the transition region contains no integration points, then steep gradients of the phase field avoid penalization in the variational objective, meaning that Eq. \eqref{objhatmc} inaccurately approximates Eq. \eqref{objhat}. The possibility of such a scenario is observed numerically in Figure \ref{small_batch}, which shows the results of minimizing Eq. \eqref{objhatmc} with 5000 epochs of ADAM and a reduced integration batch of $B=500$. This smaller batch size allows the minimizer of the approximate objective to contain larger gradients in the phase field, as indicated by the sharpness of the transition. 

\begin{figure}[hbt!]
\centering
\includegraphics[width=0.99\textwidth]{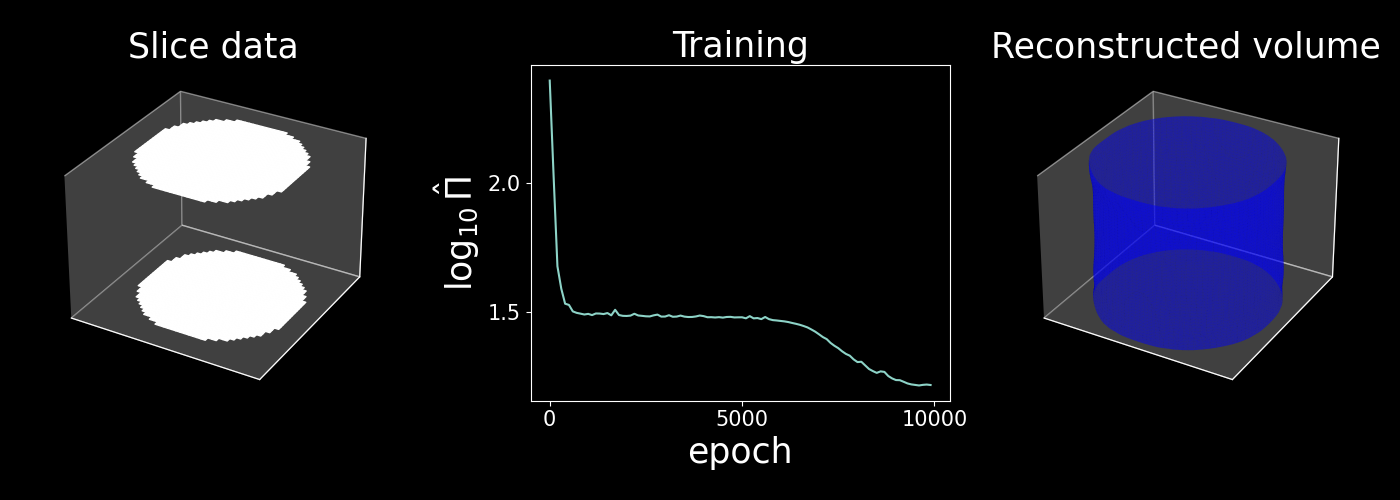}
\caption{When a fixed integration grid is used, ADAM optimization requires additional steps to converge. As in the case of the stochastic objective, a cylindrical geometry is obtained.}
\label{true_var_loss}
\end{figure}

\begin{figure}[hbt!]
\centering
\includegraphics[width=0.99\textwidth]{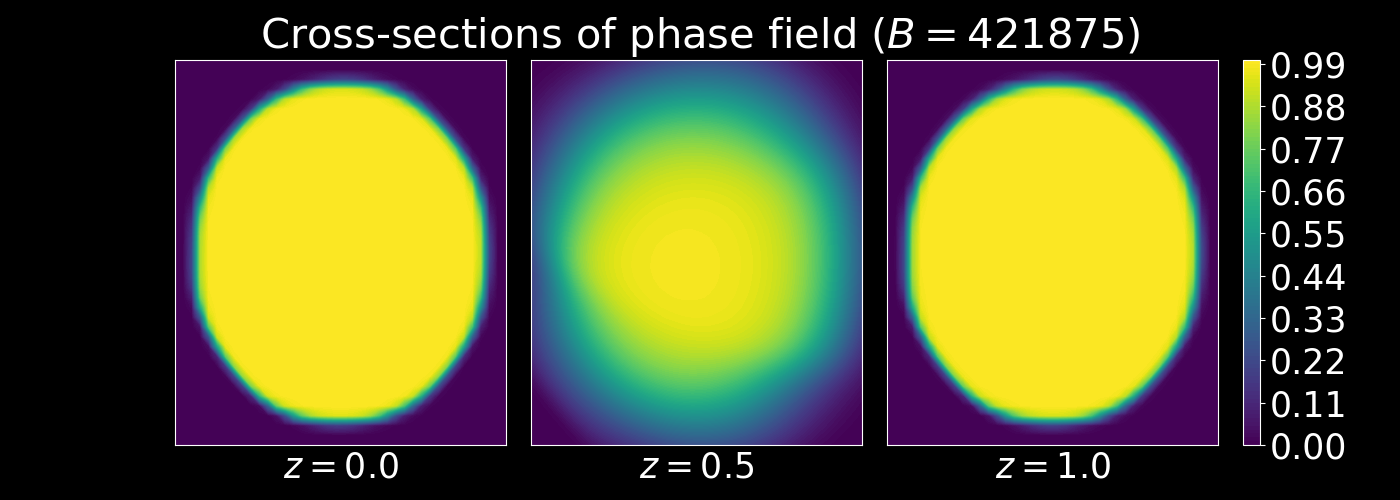}
\caption{Though the boundary surface remains cylindrical, the transition region is significantly more diffuse in the case of the dense quadrature solution. This indicates that, though it leads to high-quality reconstructed volumes, the Monte Carlo approximation of the objective is not accurate. This is not a cause for concern, given that the variational objective is simply a means to the end of obtaining reconstructions.}
\label{true_var_slice}
\end{figure}

\paragraph{} As a result of the definition of the boundary surface given by Eq. \eqref{geos}, the width of the transition region need not have a significant effect on the reconstructed geometry. It does, however, cast light on the accuracy of the Monte Carlo approximation of the variational objective. In order to further investigate the difference between the solution obtained by our stochastic approximation and the minimizer of Eq. \eqref{objhat}, we compare against a solution obtained with dense deterministic integration of the volume. In particular, we use a tensor product of 1D integration grids of 75 uniformly spaced points. This correspond to $B=421875$ total integration points. Note that Eq. \eqref{objhatmc} remains unchanged when we switch to integration on a fixed grid. In addition to increasing the size of the integration grid, we run ADAM optimization for 10000 epochs to ensure convergence. We again take $S=2$ slices with $N=1600$ noiseless measurements, and keep the penalty parameter and out-of-plane diffusion at $p=1000$ and $\epsilon_z=10$ respectively. Our goal is to compare the solution obtained with stochastic integration to an approximation of the minimizer of the continuum variational problem. See Figure \ref{true_var_loss} for the training convergence and the reconstructed volume. As before, the reconstructed geometry is cylindrical, though we note that there is now no narrowing of the cross-section in the center of the domain. The objective takes approximately twice as many epochs to converge when integrated on a fixed grid. The run time of this problem is 30 minutes, which is 110 times longer than when the objective was approximated with Monte Carlo integration. Furthermore, there is no benefit to this increase in computational cost, as a cylindrical geometry is reconstructed in both cases. However, Figure \ref{true_var_slice} shows that the Monte Carlo objective does not accurately approximate the dense quadrature solution of the variational objective. The transition region on the $z=0.5$ cross-section is wider when using the fixed grid integration, indicating that the comparatively sharp boundaries obtained from Monte Carlo integration are an artifact of stochastically integrating regions of the domain containing gradients of the phase field. We note that the run time for deterministic integration could be alleviated with a standard finite element discretization of the problem, however this example demonstrates that the approximate variational solution is preferable to the dense quadrature solution. 

\paragraph{} We propose that the quality of the reconstructed volume is of more interest than exactly solving the variational problem. Thus, though our proposed Monte Carlo integration scheme with batches of size $B=5000$-$10000$ may inaccurately compute integrals of the modified Cahn-Hilliard energy, this is not a problem so long as the reconstructed volumes are satisfactory. Despite the error incurred from the sparse random integration, the following examples will convince the reader of the efficacy of our proposed method.

\subsection{Sideways cylinder}

\paragraph{} We now test a geometry whose axis is not aligned with the slice direction. Namely, we take the same cylindrical geometry and rotate it so that the circular cross-sections lie in the $z$-$y$ plane. The level set is given by

\begin{equation*}
    \Phi(x,y,z) = (z-0.5)^2+(y-0.5)^2-0.2.
\end{equation*}

\begin{figure}[hbt!]
\centering
\includegraphics[width=0.99\textwidth]{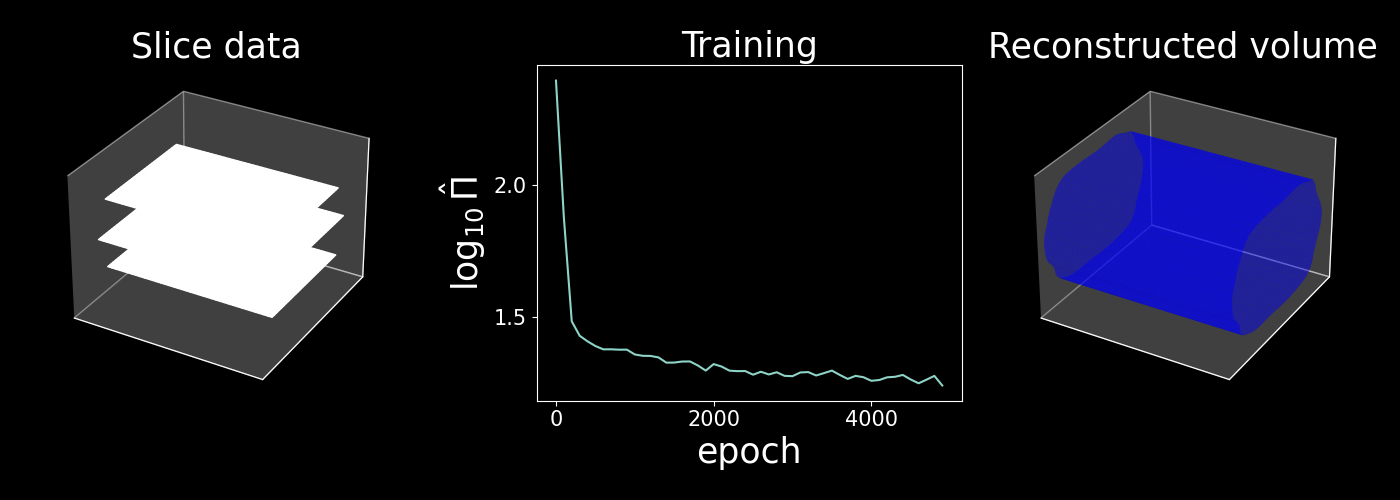}
\caption{Reconstruction of a cylinder from 5 slice planes which are perpendicular to its axis. Our method recovers the circular cross-sections in the $y$-$z$ from the rectangular slices in the $x$-$y$ plane.}
\label{sideways}
\end{figure}

\begin{figure}[hbt!]
\centering
\includegraphics[width=0.99\textwidth]{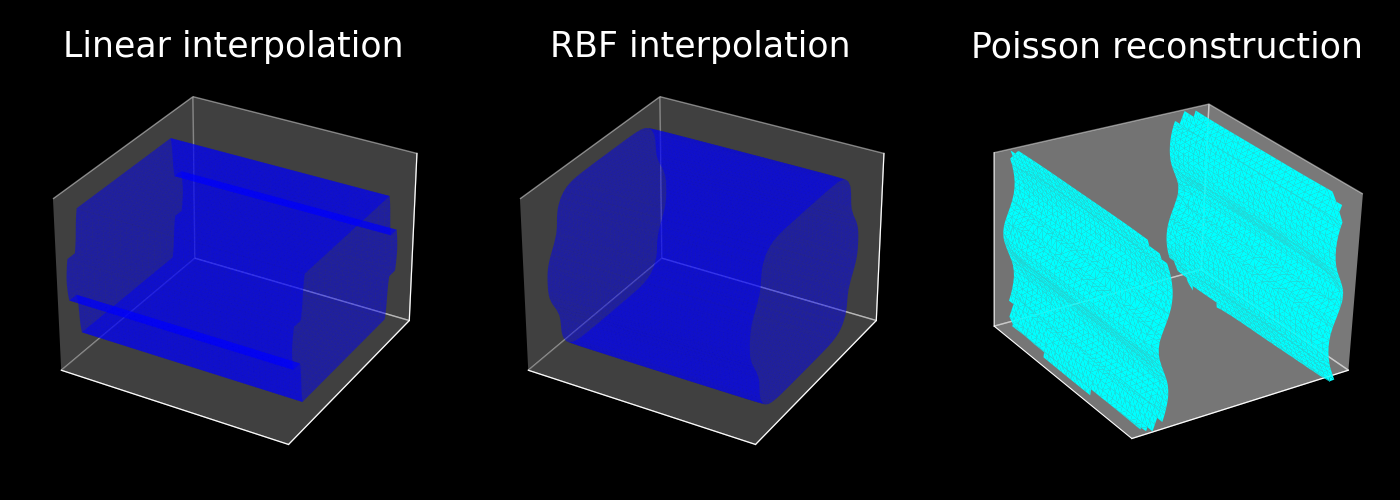}
\caption{Standard interpolation schemes without regularization do not satisfactorily interpolate the sparse slice data. Linear interpolation produces sharp corners on the reconstructed surface, and the RBF interpolation produces an undesirably ``boxy'' geometry. As a result of the sparsity of the slices, the Poisson reconstruction does not yield cross-sections with closed boundaries.}
\label{sideways_alternative}
\end{figure}

We take $N=1600$ noiseless measurements in each of $S=5$ slice planes. The penalty parameter is $p=1000$, the diffusion in the slice direction is $\epsilon_z=2.5$, the integration batch size is $B=5000$, and we run ADAM for $5000$ epochs. The results of the volume reconstruction are shown in Figure \ref{sideways}. As desired, an approximately cylindrical geometry is recovered in spite of never explicitly observing circular cross-sections. If we round the phase field with a threshold of $0.5$ to classify the phase of the material, we find that all the points in the slice planes take on the correct phase. Figure \ref{sideways_alternative} shows the reconstructed volume from the PyTorch $\texttt{torch.nn.functional.interpolate}$ linear interpolation, the $\texttt{scipy.interpolate.RBFinterpolator}$ radial basis function (RBF) interpolation, and the Poisson surface reconstruction \cite{kazhdan_poisson_2006}. We use the same $5$ slices and provide no regularization. We note that the Poisson reconstruction requires normals to the surface, which we compute with the gradient of the level set in the slice planes at boundary points. We use the ``open3d'' implementation of Poisson reconstruction in Python, which explicitly returns a triangulation of the surface. Given the sparsity of the slices, the linear interpolation leads to a boxy surface with sharp corners. Similarly, the RBF interpolation leads to cross-sections with more square features than the neural network interpolation regularized with our anisotropic energy functional. Unlike the other two interpolation methods, the Poisson reconstruction does not even provide closed cross-sections. This suggests that the combination of the neural network discretization and the Deep Ritz energy functional provide regularization that gives rise to desirable features of reconstructed volumes, such as smooth, closed surfaces free of localized high-curvature features.

\begin{figure}[hbt!]
\centering
\includegraphics[width=0.99\textwidth]{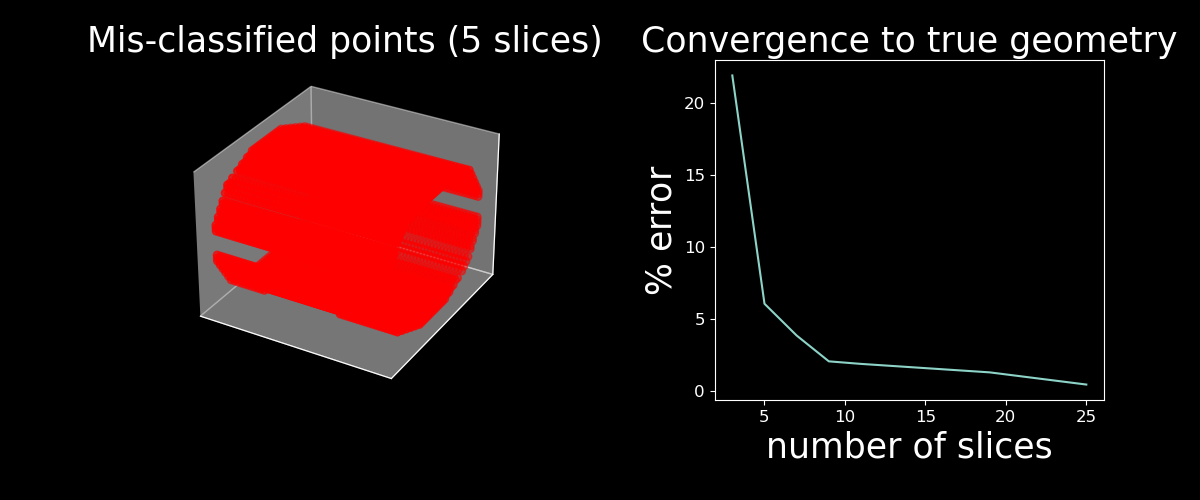}
\caption{Convergence of the network to the true geometry as the number of slices is increased. We show the set of mis-classified points for the sparse slice network. The error is near the boundaries, where the reconstruction relies on inductive biases of the network to determine the boundary shape rather than data.}
\label{sideways_convergence}
\end{figure}

\paragraph{} We also study the convergence of the reconstructed volume to the true geometry by varying the number of slices and computing the error with the true geometry. We generate a $50 \times 50 \times 50$ background grid, use the level set to assign the true phase to each grid point, evaluate the network on the same grid, round to obtain the phase classification, then compute the percent of points that are mis-classified. This is shown in Figure \ref{sideways_convergence}. The network converges to small $\%$ error as the number of slices is increased. We note that in this example and the following ones, the reconstructed geometry does not exhibit strong dependence on the initialization of the network.

\subsection{Two-way branching geometry}

\paragraph{} Thus far, we have not made use of the method to treat noisy measurement data introduced in Section 2. In this example, we reconstruct a branching blood vessel geometry from sparse slices in the presence of measurement noise. The noise level is controlled by a parameter $\sigma$, as described below. Noise is introduced in two ways: uniformly across the slice planes to model general measurement uncertainty, and in a narrow region surrounding the boundary to represent ambiguity in the location of the phase transition. To define this blurry boundary region, introduce the "inner" and "outer" level sets $\Phi^I$ and $\Phi^O$ for the branching geometry:

\begin{equation}\label{IO}
\begin{aligned}
    \Phi^I(x,y,z) = \Big( \cos 2\pi x - (1-2z)\Big)^2 + 9(y-0.5)^2 - 0.2 , \\
    \Phi^O(x,y,z) = \Big( \cos 2\pi x - (1-2z)\Big)^2 + 9(y-0.5)^2 - 0.5 .
\end{aligned}
\end{equation}

The thickness of the ambiguous boundary region is set by the constants $0.2$ and $0.5$ in Eqs. \eqref{IO}. In each slice plane, the noisy phase values of the material are given by

\begin{equation}\label{noisy_phase}
    \phi(\mathcal{S}_i) = \begin{cases} (1-\sigma) + \mathcal{U}(0,\sigma), \quad \Phi^O(\mathcal{S}_i) , \Phi^I(\mathcal{S}_i) < 0,\\
    \mathcal{U}(0,\sigma), \quad \Phi^O(\mathcal{S}_i) , \Phi^I(\mathcal{S}_i) > 0,\\
    \mathcal{U}(0,1), \quad \Phi^O(\mathcal{S}_i) < 0, \Phi^I(\mathcal{S}_i) > 0,
    \end{cases} \quad i=1,2,\dots,S.
\end{equation}

This construction yields three regions: the interior, where the true phase is near 1 but is perturbed on one side by a uniform random value in $[0, \sigma]$; the exterior, where the phase is near 0 but perturbed in the same way; and a transition region, where no phase value is assumed, and the measurement is fully ambiguous. This formulation reflects realistic imaging scenarios where uncertainty is greatest near the boundaries of the volume.

\begin{figure}[hbt!]
\centering
\includegraphics[width=0.99\textwidth]{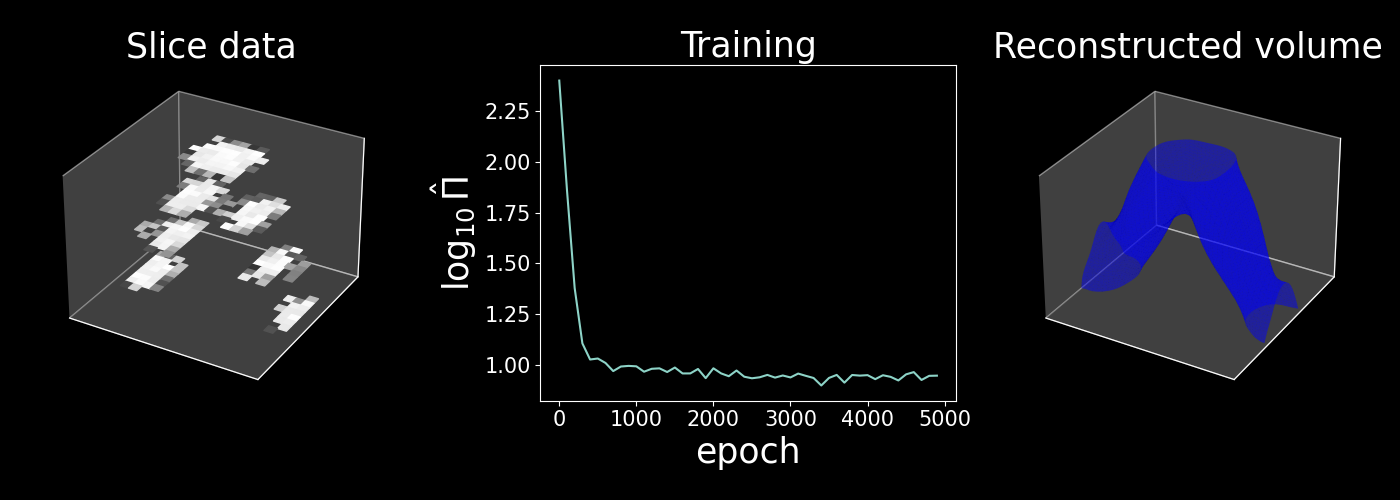}
\caption{Given the importance of volume reconstruction in biomedical imaging, we choose a level set geometry that mimics a branching blood vessel. Despite the sparse slice data and measurement noise, the branching geometry is accurately reconstructed.}
\label{branching}
\end{figure}

\paragraph{} In this example, we set the noise magnitude to be $\sigma=0.1$ and, per Eq. \eqref{sets}, the rounding cutoff to be $c=0.75$. We use $S=4$ slice planes and sparsify the number of in-plane measurements by taking $N=400$. The remaining problem parameters are set at $\epsilon_z=5$, $p=1000$, $B=5000$, and run 5000 steps of ADAM optimization. See Figure \ref{branching} for the results. Despite the sparse slice data, the increased sparsity of in-plane measurements, the presence of noise, and the complexity of the geometry, the minimizer of the stochastic objective gives rise to a smooth boundary surface which reproduces the qualitative features of the branching geometry. This problem has a run time of 24.9 seconds. Like the previous example, if all of the phase labels are rounded---both from the noisy data and the reconstructed volume---the network correctly classifies all the slice data.

\begin{figure}[hbt!]
\centering
\includegraphics[width=0.95\textwidth]{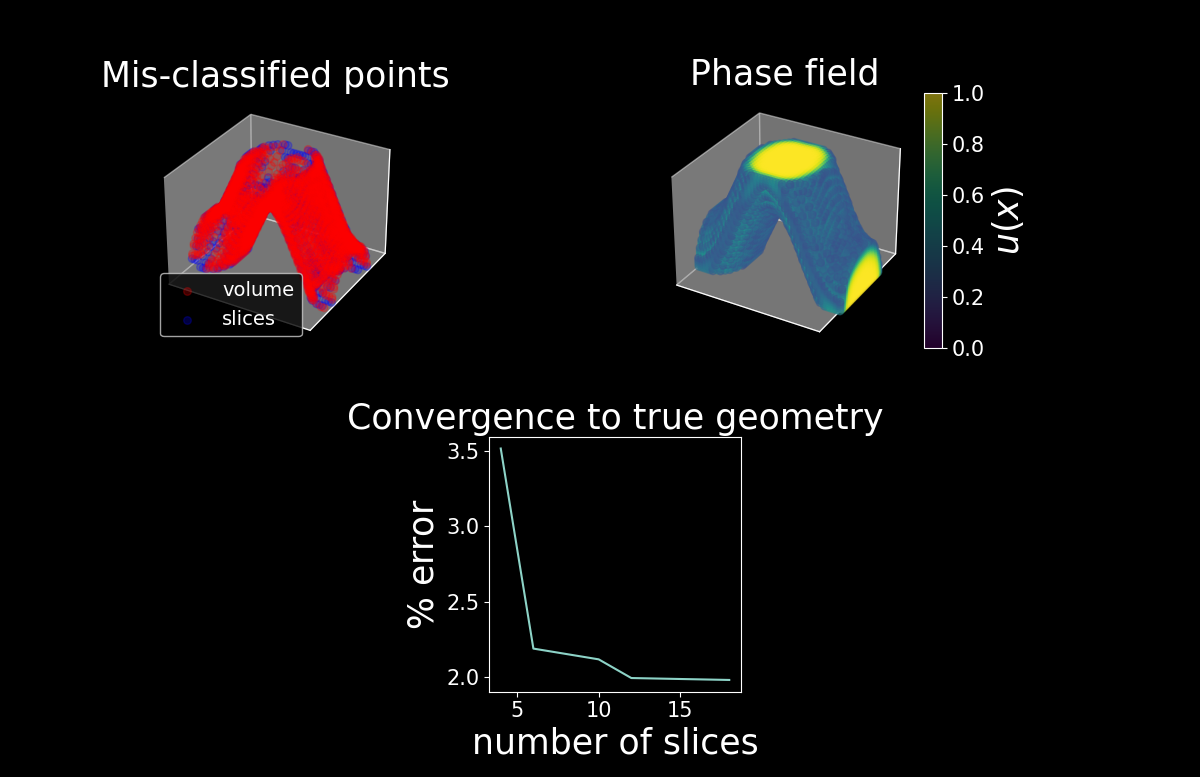}
\caption{When the the resolution of the slice plane data increases, the phase field mis-classifies points in these planes because the transition region cannot be made arbitrarily sharp. Here, we show the mis-classified points from the slices and $50\times50\times50$ integration grid for a solution generated from $18$ slices, as well as the corresponding phase field. The error with the true geometry saturates at around $2 \%$ given the diffuse boundary region.}
\label{branching_convergence}
\end{figure}

\paragraph{} We also perform a study of the convergence of our method to the true geometry as the number of slice planes increases. Because we do not expect to recover the true geometry in the presence of a noisy boundary region, we switch to deterministic data given by the outer level set $\Phi^O$ of Eq. \eqref{IO}. We increase the number of measurements in the slice planes to $N=2500$ in order to not limit convergence by the coarseness of the data. All other problem parameters remain the same. See Figure \ref{branching_convergence} for the results of the convergence study as well as a visualization of the mis-classified data. We note that when the the phase field has a transition region of non-zero thickness. This transition region causes a thin shell of points near the boundary to be mis-classified, even as the number of slice planes increases. Increasing the number of slice planes does not drive the error with the true geometry below $2\%$. However, we note that increasing the penalty parameter to $p=5000$ causes the reconstructed geometry to obtain $1\%$ error with $18$ slices. Evidently, the diffusion term is regularizing the phase field to some extent, as de-emphasizing it in the objective leads to a sharper boundary. 

\subsection{Four-way branching geometry}

\paragraph{} To test our method with increased geometric complexity and noise, we study a four-way branching geometry with significant measurement noise. We set the noise magnitude to be $\sigma=0.3$, take $S=5$ slices, and consider $N=625$ measurement points per slice. The four-way branch is represented by inner and outer level sets given by:

\begin{equation*}
\begin{aligned}
    \Phi^I(x,y,z) = \Big( \cos 2\pi x - (1-2z^3)\Big)^2 + 9(y-0.5)^2 
    \\+ \Big( \cos 2\pi y - (1-2z^3)\Big)^2 + 9(x-0.5)^2  - 2.75, \\
    \Phi^O(x,y,z) = \Big( \cos 2\pi x - (1-2z^3)\Big)^2 + 9(y-0.5)^2 
    \\+ \Big( \cos 2\pi y - (1-2z^3)\Big)^2 + 9(x-0.5)^2 - 3.25 .
\end{aligned}
\end{equation*}

\begin{figure}[hbt!]
\centering
\includegraphics[width=0.99\textwidth]{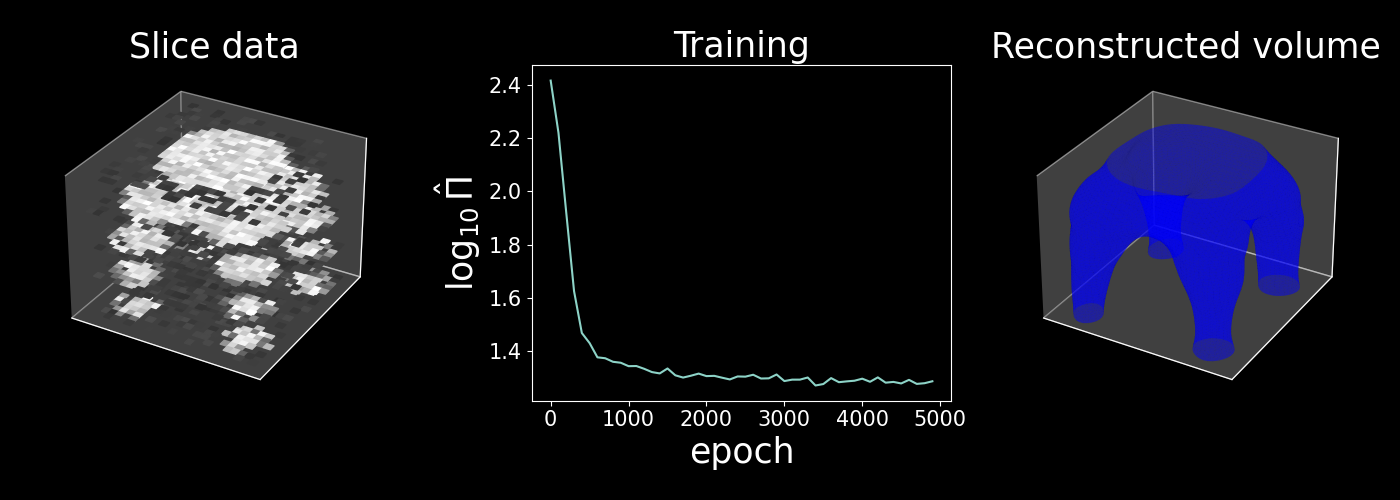}
\caption{As a result of the noise in the image, 831 measurement points are considered to have ambiguous phase and are thus not penalized in the regression loss. Nonetheless, our reconstructed volume has smooth boundary surface and reproduces qualitative features of the slice data.}
\label{dinosaur}
\end{figure}

Using these two level sets, Eq. \eqref{noisy_phase} is used to generate the noisy phase data. The integration batch size is $B=5000$, the penalty parameter is $p=1000$, and the out-of-plane diffusion is $\epsilon_z=5$. See Figure \ref{dinosaur} for the results. This problem has a run time of 26.4 seconds. Per our definition of the inside and outside phase measurement points in Eq. \eqref{sets}, there are 575 points assigned to the inside phase and 1719 points assigned to the outside phase. This leaves 831 measurement points not assigned as a result of noise in the slice images. Despite the ambiguous phase labels of more than one third of the points, the reconstructed volume is smooth and obtains the important qualitative features of having four cylindrical legs and no disconnected regions. Once again, when phases are assigned with rounding, the network correctly classifies all the slice data.

\subsection{Hollow and tilted cylinder}

\paragraph{} This test case considers a volume with different topology than the previous examples. We study a hollow cylinder which is tilted such that the cross-sections are offset from each other in the $x$ direction. The hollow interior introduces a topological feature absent in previous examples, while the tilt poses a new challenge for interpolation between sparse slices. We omit measurement noise in order to isolate geometric complexity as the variable of interest. The level set which represents the hollow tilted cylinder is

\begin{equation*}
\begin{aligned}
    \Phi(x,y,z) = \Big( ( x - z/5 - 0.4)^2 + (y-0.5)^2 - 0.15 \Big) 
    \\ \Big( (x-z/5-0.4)^2 + (y-0.5)^2 - 0.05\Big).
\end{aligned}
\end{equation*}

\begin{figure}[hbt!]
\centering
\includegraphics[width=0.99\textwidth]{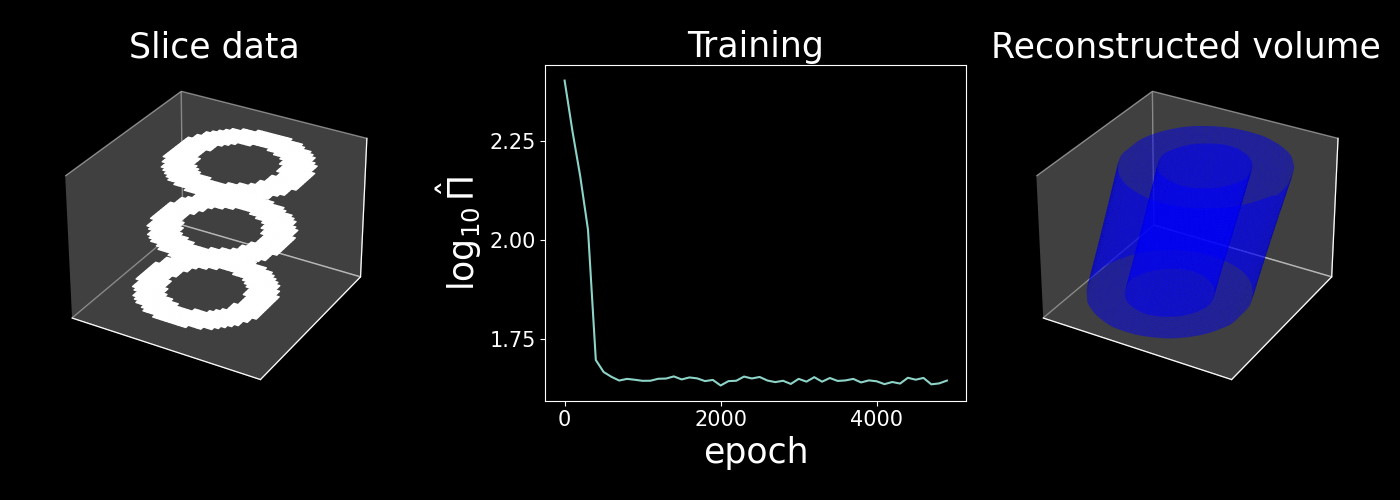}
\caption{Constructing a tilted cylinder from three noiseless slices. The outer surface of the reconstructed geometry linearly interpolates the offset cross-sections, but this plot does not give insight into the inner boundary of the hollow cylinder.}
\label{donut}
\end{figure}

\begin{figure}[hbt!]
\centering
\includegraphics[width=0.99\textwidth]{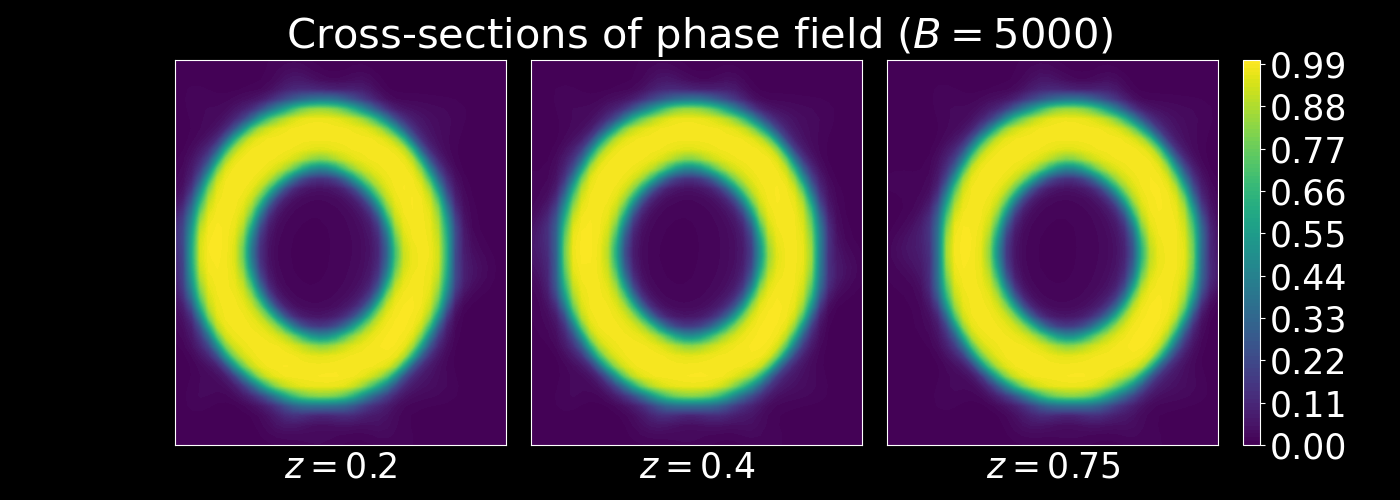}
\caption{The phase field on cross-sections of the reconstructed geometry which do not contain measurement data. Our method reproduces the topology of the hollow cylinder, interpolates the offset cross-sections, and obtains sharp boundaries. The variational objective fits cylindrical surfaces which are parallel to one another.}
\label{donut_x}
\end{figure}

\begin{figure}[hbt!]
\centering
\includegraphics[width=0.99\textwidth]{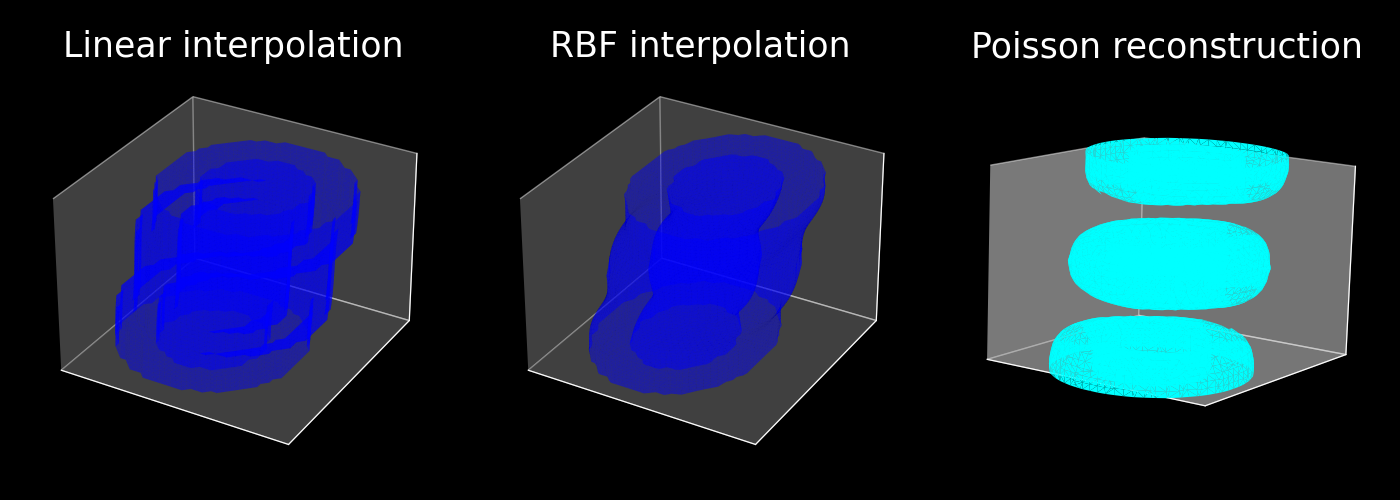}
\caption{The surface reconstructed from sparse slice data using standard interpolation schemes in PyTorch and SciPy has a number of unappealing features. The linear interpolation produces a surface with sharp corners, whereas the RBF interpolation does not reproduce the straight sides of the tilted cylinder. Due to the sparsity of the slices, the Poisson reconstruction yields three disconnected regions.}
\label{donut_alternative}
\end{figure}

Given the thin features of the geometry to reconstruct, we increase the neural network width to 50 units per layer. We take $S=3$ slice planes, $N=1600$ measurement points per plane, a penalty parameter of $p=1000$, an integration batch size of $B=5000$, and run ADAM optimization for 5000 epochs. We set the out-of-plane diffusion to $\epsilon_z=1$ (isotropic diffusion), as we anticipate $z$ gradients of the phase field given that the slice planes are offset from each other. See Figure \ref{donut} for the slice data, the training convergence, and the reconstructed geometry. The phase field converges quickly to a cylinder whose outer surface linearly interpolates the offset slice data. In order to visualize the interior of the geometry, we show heat maps of three cross-sectional slices which do not contain measurement data in Figure \ref{donut_x}. We recover the hollow cylindrical geometry and note that the inner cylindrical surface is parallel to the outer cylindrical surface, even away from the slice planes. Contrast this with results from Pytorch's linear interpolation, SciPy's RBF interpolator, and the open3d implementation of Poisson reconstruction. Using the same configuration of slice data, the surfaces reconstructed from these three methods are shown in Figure \ref{donut_alternative}. The linear interpolation produces vertical walls of the cylinder whose position shifts discontinuously at the slice planes. The RBF interpolation performs better, but does not produce straight-sided cylinder walls as the Deep Ritz interpolation does. The Poisson reconstruction  produces three disconnected surfaces, much like the hourglass geometry with insufficient out-of-plane diffusion seen in Figure \ref{sweep}. With our method, a connected cylindrical geometry with straight sides and circular cross-sections is obtained in a run time of 27.1 seconds.

\subsection{Stanford Bunny}

\begin{figure}[hbt!]
\centering
\includegraphics[width=0.99\textwidth]{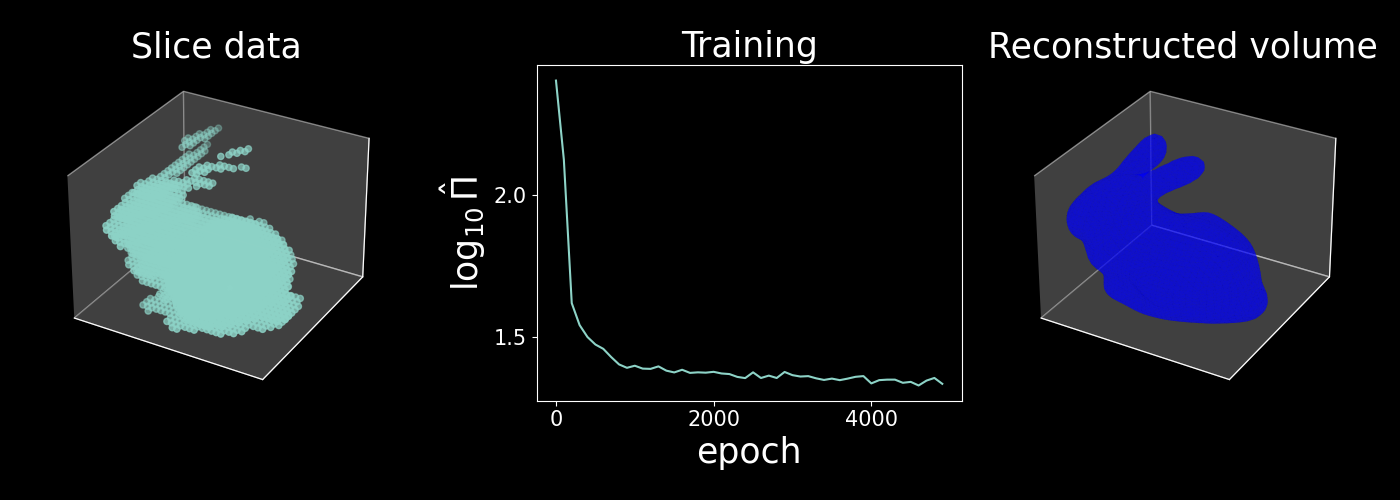}
\caption{Reconstructing the Stanford Bunny from slice data. The slice data is shown here as a scatter plot of the interior points to facilitate visualization.}
\label{bunny}
\end{figure}

\begin{figure}[hbt!]
\centering
\includegraphics[width=0.99\textwidth]{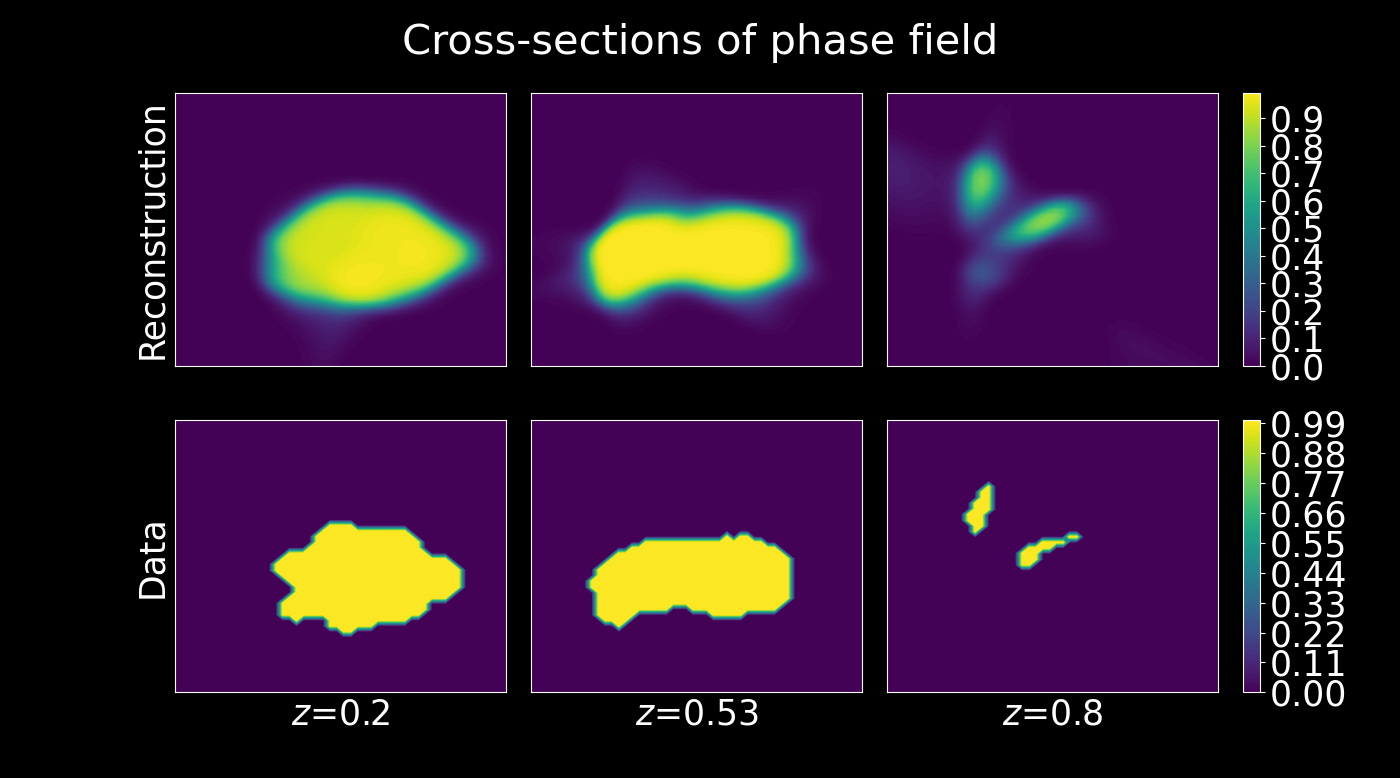}
\caption{Comparing the phase field and the measurement data on three slice planes. The phase field diffuses the transition between phases and smooths some small-scale features of the boundary surface, but manages to represent qualitative features of the geometry very well, including the tips of the bunny's ears.}
\label{bunny_comparison}
\end{figure}

\paragraph{} As a more challenging validation of our method, we reconstruct the Stanford Bunny from slice data---a well-known benchmark in computer graphics and 3D shape reconstruction. Owing to the complexity of the geometry, we maintain the width of the hidden layers at $50$ and take $S=15$ slice planes with $N=2500$ noiseless measurements points per plane. The penalty parameter is set at $p=1000$, the integration batch size is increased to $B=7500$, and we again run $5000$ epochs of ADAM optimization. The out-of-plane diffusion is set at $\epsilon_z=2.5$. See Figure \ref{bunny} for the slice data, the training convergence, and the reconstructed geometry. As in previous examples, the variational objective converges quickly and the reconstructed geometry contains the salient qualitative features of the measurement data. The run time of this problem increases to 166.7 seconds, which is a consequence of the larger integration batch size and set of measurement data. In order to better understand the performance of the reconstructed volume, we compare cross-sections of the phase field to the measurement data at $3$ separate slice planes in Figure \ref{bunny_comparison}. As expected, the phase field blurs the interface between phases, yet it captures complex, small-scale features of the geometry such as the ears of the bunny.

\begin{figure}[hbt!]
\centering
\includegraphics[width=0.99\textwidth]{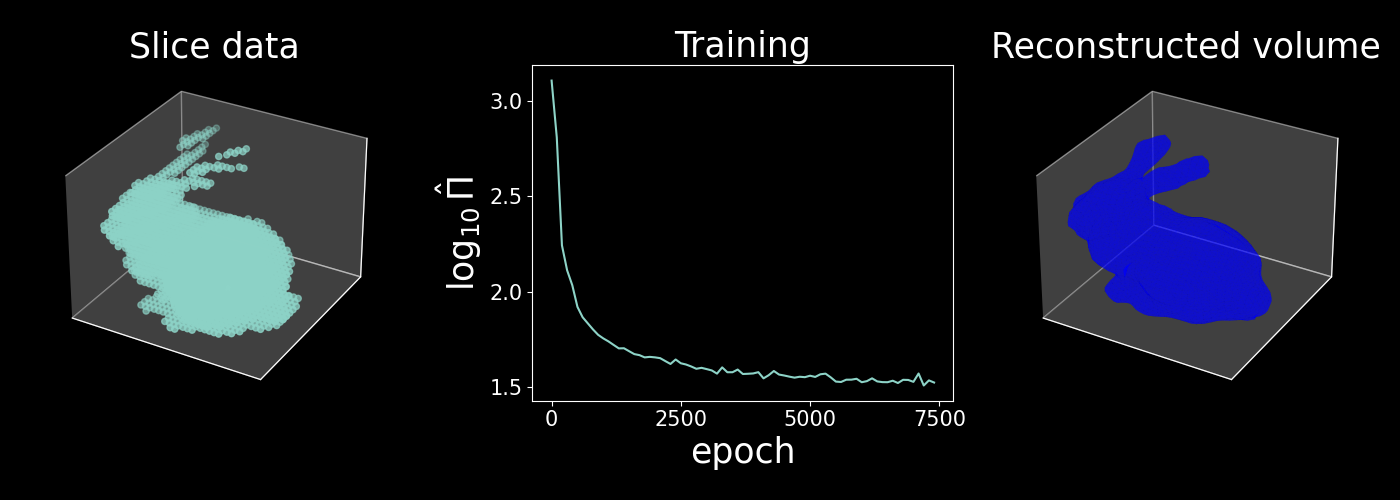}
\caption{A more expressive network with a larger penalty on the measurement data and smaller diffusion parameters leads to a higher-fidelity reconstruction. Such modifications to the problem parameters are advisable when the geometry to be reconstructed has small-scale features.}
\label{bunny_2}
\end{figure}

\begin{figure}[hbt!]
\centering
\includegraphics[width=0.99\textwidth]{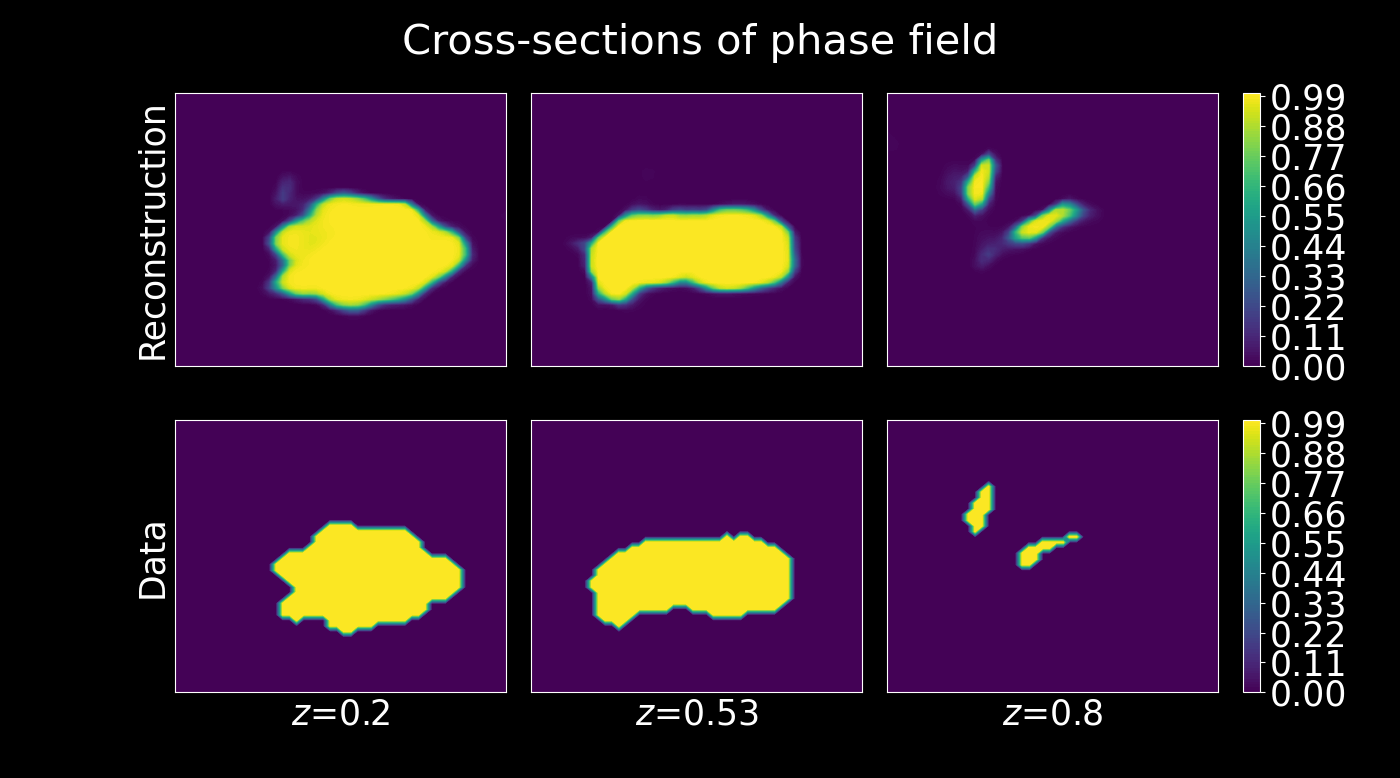}
\caption{Comparing the phase field and the measurement data on three slice planes with updated problem parameters. The width and smoothness of the interface in slices can be adjusted with our method by tuning the penalty on the slice data and the diffusion parameters. }
\label{x_2}
\end{figure}

\paragraph{} We note that the features of the bunny seen in Figure \ref{bunny_comparison} are not as sharp as one might desire. The previous examples have shown that the tendency of the proposed method to find smooth geometries has been a useful ``inductive bias'' to interpolate the slice data. In the case of the Stanford Bunny, the cross-sectional slices themselves have many small scale features that are not accurately captured by the reconstruction with the current problem parameters. To show that our method can be tuned to represent these small scale features in the slices, we re-run the problem with a larger penalty ($p=5000$), smaller diffusion ($\boldsymbol \epsilon=[0.5,0.5,1]$), a neural network with the width increased to $75$ hidden units, an integration batch size of $B=7500$, and $7500$ steps of ADAM optimization. We expect that smaller diffusion and a larger penalty will improve the fidelity of the reconstruction within slice planes. See Figures \ref{bunny_2} and \ref{x_2} for the results. We notice additional fine-grained features of the bunny showing up in the reconstructed volume, as well as sharper interfaces which more closely follow the slice data. The slices of the reconstruction more closely resemble the original measurement data with these changes to the problem parameters.

\subsection{Biomedical examples}

\begin{figure}[hbt!]
\centering
\includegraphics[width=0.25\textwidth]{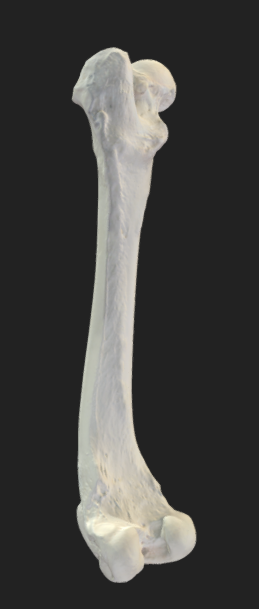}
\caption{Our first biomedical test case is the femur bone of a dog. The geometry is reconstructed from $30$ slice planes. "Femur (os femoris) dog" (https://skfb.ly/6TLF8) by vetanatMunich is licensed under CC Attribution-NonCommercial-NoDerivs (http://creativecommons.org/licenses/by-nc-nd/4.0/).}
\label{dog_truth}
\end{figure}

\begin{figure}[hbt!]
\centering
\includegraphics[width=0.99\textwidth]{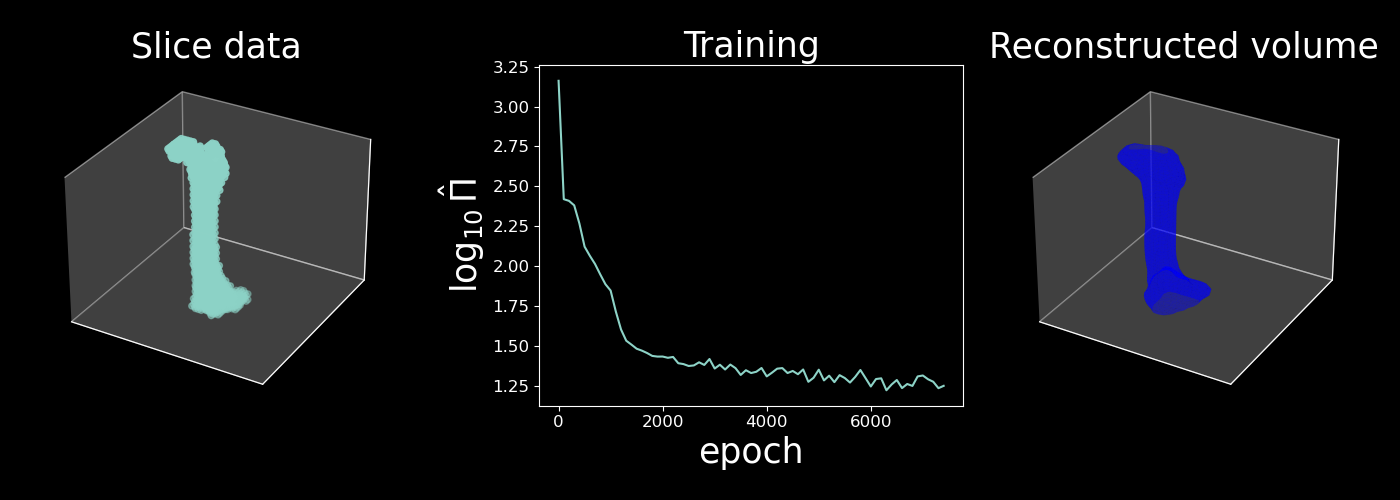}
\caption{We reconstruct a femur bone using $30$ slice planes. This and the subsequent example are taken from biomedical imaging applications to showcase the efficacy of our method in this context.}
\label{dog_reconstruct}
\end{figure}

\begin{figure}[hbt!]
\centering
\includegraphics[width=0.99\textwidth]{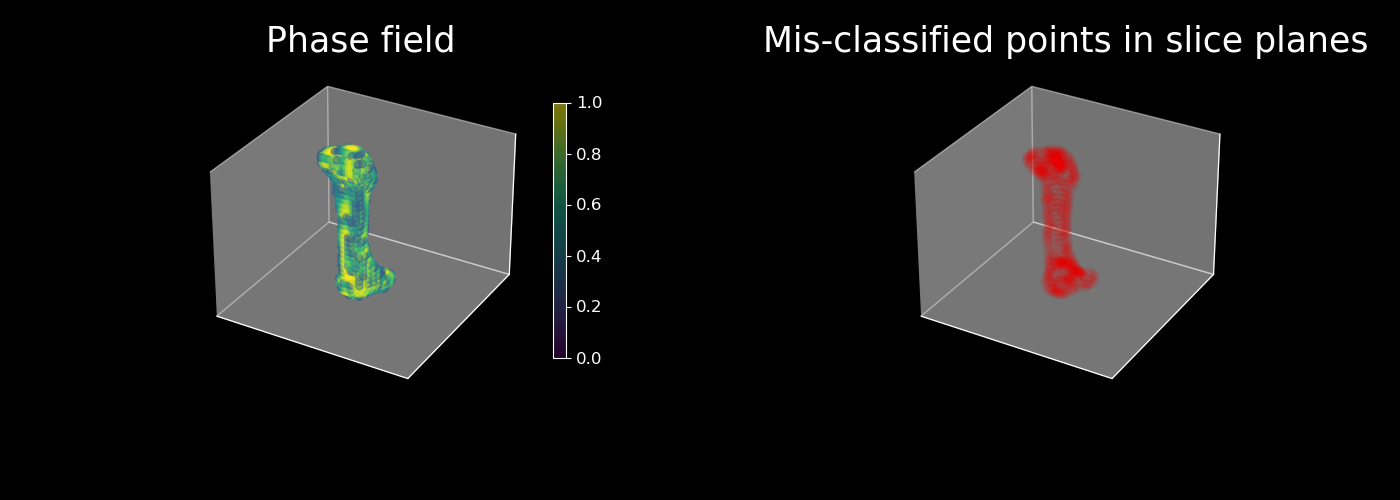}
\caption{The mis-classified points from slice planes are all near the surface of the volume, as the neural network discretization has to effectively represent a discontinuity in the phase field at these points. The percent of mis-classified points is $0.5\%$.}
\label{dog_error}
\end{figure}

\paragraph{} As a final validation of our method, we reconstruct two volumes motivated by real biomedical imaging applications. The first comes from a 3D scan of the femur bone of a dog, which is shown in Figure \ref{dog_truth}. The slice data used in the reconstruction problem comprises $30$ slice planes, each with $80 \times 80=6400$ evenly-spaced points. As in the case of the Stanford bunny, the settings for our reconstruction method are $\boldsymbol \epsilon=[0.5,0.5,1]$ and $B=7500$, but we up the penalty to $p=7500$. ADAM optimization is run for $7500$ epochs and the network has a width of $75$. See Figure \ref{dog_reconstruct} for the results. Keeping the same hyperparameters as the previous example, our method returns an accurate reconstruction of this volume. Using the error measure introduced previously, we find that the fraction of mis-classified points is $0.5\%$. See Figure \ref{dog_error} for a characterization of the mis-classified points.

\begin{figure}[hbt!]
\centering
\includegraphics[width=0.6\textwidth]{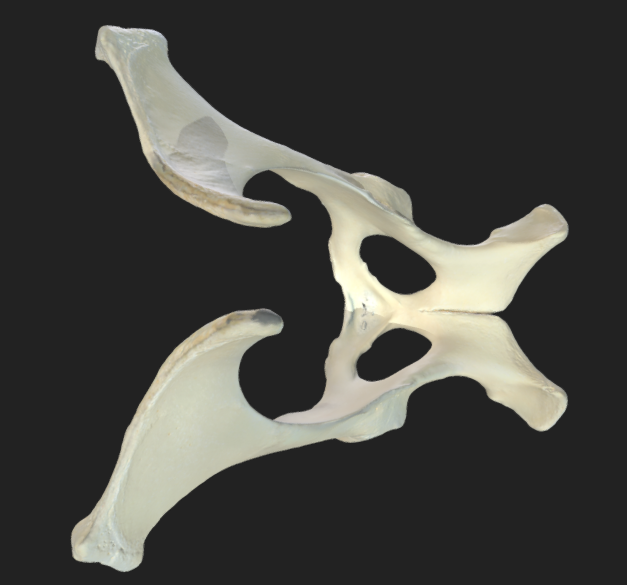}
\caption{The ground truth volume of the pelvic bone of a donkey. We use our method to reconstruct this volume from $35$ slice planes. ``Pelvic bones (ossa coxarum) donkey'' (https://skfb.ly/6SIpQ) by vetanatMunich is licensed under CC Attribution-NonCommercial-NoDerivs (http://creativecommons.org/licenses/by-nc-nd/4.0/).}
\label{donkey_truth}
\end{figure}

\begin{figure}[hbt!]
\centering
\includegraphics[width=0.99\textwidth]{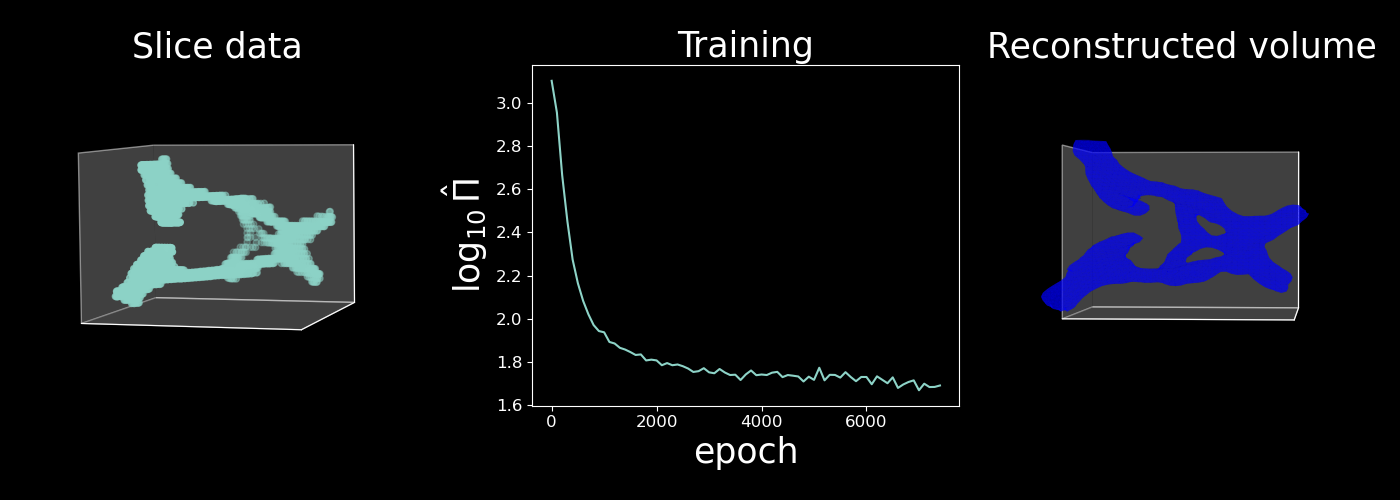}
\caption{Our proposed method accurately reconstructs the complex geometry of the pelvic bone from $35$ planes. Note that this is the first geometry we test with two holes in it. Furthermore, we re-used the hyperparameter settings of the previous example to obtain this result, meaning no tuning was necessary.}
\label{donkey_reconstruct}
\end{figure}

\paragraph{} The second biomedical example is the pelvic bone of a donkey, shown in Figure \ref{donkey_truth}. The geometry is obtained from a 3D scan of the bone, and the slice data used in the reconstruction problem comprises $35$ slice planes, each with $80 \times 80=6400$ evenly-spaced points. We use exactly the same hyperparameter settings as the femur example. See Figure \ref{donkey_reconstruct} for the results. The qualitative agreement of the reconstructed volume with the ground truth in Figure \ref{donkey_truth} is very strong. The curving flanges on the left side are accurately reconstructed, as well as the two holes in the center right. Fine-grained details, such as the texture of the geometry and the thin crack between the two central holes are not recovered by the reconstructed volume, but these features do not appear in the slice data either. The percent of mis-classified points is $2.2\%$, again all near the surface. In order to drive the error down to zero, the network must represent a field which is discontinuous on the surface of the complex surface defined by the pelvic bone. As such, we do not find it surprising that the percent error is larger with the more complex geometry. Of course, another source of error is the limited number of slices, which obscures fine features of the original geometry. Despite the smattering of mis-classified points, the phase field obtained with our method leads to a satisfactory reconstruction of the boundary surface.

\section{Conclusion}

\paragraph{} In past model-based volume reconstruction works, reliance on complex preprocessing steps, dense slice data, and mesh-based implementations has presented obstacles to devising robust SVR methodologies for clinical use. We have introduced a volume reconstruction method which avoids image segmentation by operating on noisy slice data directly, is able to reconstruct volumes from widely spaced slices, and converges to a solution in a matter of seconds. We view the variational objective, stochastic integration, the neural representation, and gradient-based optimization all coming together to induce a geometric prior on reconstructed shapes. We emphasize that, in the sparse-slice regime, success should not be interpreted as exact recovery of a uniquely defined volume away from the measured planes. Instead, the relevant criterion is whether the method produces reconstructions that honor the observed slices and behave plausibly in the unobserved regions, and our numerical examples suggest that the proposed method does so consistently, including on complex biomedical geometries. After tuning free parameters of the model, we demonstrated its ability to handle wide slice spacing, measurement noise, geometric complexity, and topological changes with a number of examples. We discussed an example of discrepancies between the dense-grid quadrature minimizer of the variational objective and solutions obtained by the Monte Carlo integration. In the sparse-slice regime, reconstruction quality is therefore assessed not only through agreement with the observed slices, but also through geometric behavior in unobserved regions, including smoothness, connectivity, topology, and the avoidance of unsupported high-curvature features. We reiterate that our method need not accurately compute a solution to the variational problem in order to be successful. However, a reconstruction method should produce geometries that avoid high-curvature features which are not suggested by the slice data, including disconnected features between slices. In fact, our results indicate that in addition to satisfying these desiderata, our method exhibits a number of other desirable properties, such as drastic reductions in computational cost and sharper boundaries of the phase field representing the reconstructed volume. We also showed with example problems that our method performed better than the three standard reconstruction techniques of linear interpolation, RBF interpolation, and Poisson reconstruction, suggesting that the combination of variational energy, neural representation, stochastic underintegration, and nonlinear optimization provides useful regularization for these sparse reconstruction problems. Future work will include modifications of the geometric regularization in order to customize properties of the reconstructed volume. Furthermore, establishing specific guidelines for the choice of problem parameters like the data penalty, integration batch size, and diffusion tensor will be an important next step to preparing this method for use in the field. This will require in-depth studies across a number of geometries which detail how features of the reconstructed volumes vary with the problem parameters. Furthermore, while the present work focuses on reconstruction without reliance on prior datasets, the proposed variational framework is naturally extensible to incorporate additional prior knowledge when available. For example, one could augment the objective with terms that favor membership in a library of plausible geometries or enforce known structural features of the target class. Such extensions would allow the method to interpolate between purely data-driven and purely model-based reconstruction, providing a mechanism to control how strongly the reconstruction adheres to the measured slice data versus additional prior information in severely underdetermined settings. Finally, more in-depth mathematical analysis may shed light on the success of the anisotropic diffusion regularization, as well as the differences between minimizers of the continuum and stochastic variational objectives.



\end{document}